\documentclass[preprint,eqsecnum,aps,nofootinbib]{revtex4}
\usepackage{amsfonts,amsmath,amssymb,amsthm}
\usepackage{latexsym}
\usepackage{bbm,bm}
\usepackage{graphicx}

\newcommand{\ket}[1]{\lvert #1 \rangle}
\newcommand{\bra}[1]{\langle #1 \lvert}
\newcommand{\beq}{\begin{equation}}
\newcommand{\eeq}{\end{equation}}
\newcommand{\beqs}{\begin{eqnarray}}
\newcommand{\eeqs}{\end{eqnarray}}

\begin{document}

\title{Thermal Entanglement and Thermal Discord in two-qubit Heisenberg XYZ Chain with Dzyaloshinskii-Moriya Interactions}

\author{DaeKil Park$^{1,2}$}

\affiliation{$^1$Department of Electronic Engineering, Kyungnam University, Changwon
                 631-701, Korea    \\
             $^2$Department of Physics, Kyungnam University, Changwon
                  631-701, Korea    
                      }

\begin{abstract}
In order to explore the effect of external temperature $T$  in quantum correlation we compute thermal entanglement and thermal discord analytically in the 
Heisenberg $X$ $Y$ $Z$ model with  Dzyaloshinskii-Moriya Interaction term ${\bm D} \cdot \left( {\bm \sigma}_1 \times {\bm \sigma}_2 \right)$.
For the case of thermal entanglement it is shown that quantum phase transition occurs at $T = T_c$ due to sudden death phenomenon. For antiferromagnetic 
case the critical temperature $T_c$ increases with increasing $|{\bm D}|$. For ferromagnetic case, however, $T_c$ exhibits different behavior 
in the regions  $|{\bm D}| \geq |{\bm D_*}|$ and $|{\bm D}| < |{\bm D_*}|$, where ${\bm D_*}$ is particular value of ${\bm D}$. It is shown that
$T_c$ becomes zero at $|{\bm D}| = |{\bm D_*}|$. We explore the behavior of thermal discord in detail at $T \approx T_c$. For antiferromagnetic case
the external temperature makes the thermal discord exhibit exponential damping behavior, but it never reaches to exact zero. For ferromagnetic case
the thermal entanglement and thermal discord are shown to be zero simultaneously at $T_c = 0$ and  $|{\bm D}| = |{\bm D_*}|$. 
This is unique condition for simultaneous disappearance of thermal entanglement and thermal discord in this model.
\end{abstract}

\maketitle

\section{Introduction}
Recently, much attention is paid to quantum technology developed on the foundation of quantum information processing (QIP). The physical resource of QIP 
is quantum correlation such as quantum entanglement\cite{schrodinger-35,text,horodecki09} and quantum discord\cite{discord1,discord2}. Thus, 
they are at the heart in various QIP such as  quantum teleportation\cite{teleportation},
superdense coding\cite{superdense}, quantum cloning\cite{clon}, quantum cryptography\cite{cryptography,cryptography2}, quantum
metrology\cite{metro17}, and quantum computer\cite{qcomputer,qcreview}. In particular, physical realization of quantum cryptography and quantum computer seems to be accomplished in the near future\footnote{see Ref. \cite{white} and web page
https://www.computing.co.uk/ctg/news/3065541/european-union-reveals-test-projects-for-first-tranche-of-eur1bn-quantum-computing-fund.}. 

Pure quantum mechanical phenomena can occur in ideally closed system. However, real physical systems inevitably interact with their 
surroundings. Then, quantum systems undergo decoherence\cite{zurek03} and, as a result, lose their quantum properties. Thus, the environment can 
make various QIP useless due to disappearance of quantum correlation. In order to develop the quantum technology, therefore,  it is important to study 
effect of its surroundings precisely. 

Usually, decoherence significantly changes the quantum correlation. For example, decoherence makes degradation of entanglement and discord. For the case of entanglement the entanglement sudden death (ESD) occurs sometimes when the entangled multipartite quantum system is
embedded in Markovian environments\cite{markovian,yu05-1,yu06-1,yu09-1,almeida07}. This means that the entanglement is completely disentangled at finite times. 
Most typical environment is external temperature. If external temperature induces the ESD phenomenon in a system, this implies that quantum phase transition
occurs at critical temperature $T_c$. This means that quantum entanglement completely disappears at $T \geq T_c$. The purpose of this paper is to
examine how quantum entanglement and quantum discord are degraded due to external temperature and to study on the quantum phase transition in detail
by introducing the anisotropic Heisenberg $X$ $Y$ $Z$ chain system with Dzyaloshinskii-Moriya (DM) interaction\cite{dzya58,mori60}.
The quantum phase transition of a spin-$3$ Heisenberg model with DM interaction was discussed in Ref. \cite{marko17}.

Heisenberg model is a simple spin chain model, which is used to simulate many physical systems such as nuclear spins\cite{kane98}, 
quantum dots\cite{dots}, superconductor\cite{super}, and optical lattices\cite{optics}. Since spin is two-level, Heisenberg model is ideal for generation of 
qubit states. Thus, this model attracts attention recently for realization of solid-based quantum computer. In Ref. \cite{byrnes17} the DM interaction terms are 
introduced in this model due to spin-orbit couplings. The general Hamiltonian for $N$-spin Heigenberg model with DM interaction is 
\begin{equation}
\label{general-hamil}
H_N = \sum_{i=1}^{N - 1} \left[ J_x \sigma_i^x \otimes \sigma_{i+1}^x  + J_y \sigma_i^y \otimes \sigma_{i+1}^y 
+ J_z \sigma_i^z \otimes \sigma_{i+1}^z 
 + {\bm D} \cdot \left( {\bm \sigma}_i \times {\bm \sigma}_{i+1} \right) \right],
\end{equation}
where the last term is called the DM interaction arising from spin-orbit couplings\footnote{In addition to DM interaction the spin-orbit coupling induces the
second order term called $\Gamma$ tensor\cite{shek92,suhas08,marko18}. This term is neglected in this paper. This means that our paper is 
valid only in the first order correction of spin-orbit interaction.}. The real parameters $J_{\alpha} \hspace{.2cm} (\alpha = x, y, x)$
denote the symmetric exchange spin-spin interactions, ${\bm D}$ is the antisymmetric DM exchange interaction, and 
$\sigma_i^{x, y, z}$ are Pauli spin operators on the site $i$. The negative and positive $J_{\alpha} \hspace{.2cm} (\alpha = x, y, x)$ correspond to
the ferromagnetic and antiferromagnetic nature of the system respectively. If $J_x = J_y \neq J_z$, this system is called $X$ $X$ $Z$ model with DM interaction.

In this paper we study on the effect of external temperature in the entanglement and discord based on the analytic results by introducing $2$-qubit Heisenberg
model with DM-interaction. In particular, we focus on the quantum phase transition in entanglement, which occurs due to ESD phenomena. Our motivation is 
as follows. Although definitions of entanglement and discord are completely different, these are two different measures of quantum correlation. 
Thus, we guess they should exhibit similar behavior to each other. In order to confirm our conjecture we examine in detail the behavior of 
thermal discord at $T \approx T_c$. It is shown that for antiferromagnetic case ($J_{\alpha} > 0$) the temperature dependence of discord exhibits an
exponential damping behavior, but it never reaches to zero. This means that thermal discord does not completely vanish in the separable states arising 
from thermal density matrix at $T \geq T_c$. For ferromagnetic case ($J_{\alpha} < 0$) it is shown that the critical temperature $T_c$ approaches to 
zero if the DM coupling constant ${\bm D}$ approaches to a particular value ${\bm D_*}$. At ${\bm D} = {\bm D_*}$ thermal discord complete vanishes
at $T = T_c = 0$. The point  ${\bm D} = {\bm D_*}$  and $T=0$ in parameter space is a unique point, where the thermal entanglement and thermal discord
simultaneously vanish.

The paper is organized as follows. In section II we derive the thermal density matrices $\rho_Z (T)$ when $D_x = D_y = 0$ and 
$\rho_Y (T)$ when $D_x = D_z = 0$. The case of $D_y = D_z = 0$ is not presented in this paper because all calculation is very similar to the case of 
 $D_x = D_z = 0$. If two components of ${\bf D}$ are nonzero, we should rely on numerical analysis for computation of entanglement and discord. 
Thus, these cases are not considered in this paper. In section III we compute the thermal entanglement of $\rho_Z (T)$ and $\rho_Y (T)$  analytically. 
Using our analytical results we examine the critical temperature $T_c$, above which the thermal entanglement completely vanishes. 
It is shown that for antiferromagnetic case ($J_{\alpha} > 0, \hspace{.2cm} \alpha = x, y, z$) the critical temperature is determined by single equation 
for each thermal density matrix. For ferromagnetic case ($J_{\alpha} < 0, \hspace{.2cm} \alpha = x, y, z$), however, the critical 
temperature is determined by two different equations in the two separate two regions $|{\bm D}| \geq |{\bm D_*}|$ and 
$|{\bm D}| < |{\bm D_*}|$. It is shown that $T_c$ becomes zero at the boundary of these region $|{\bm D}| = |{\bm D_*}|$.
In section IV the thermal discords of $\rho_Z (T)$ and $\rho_Y (T)$ are
analytically computed. Using the analytic results we examine the behavior of thermal discords near the critical temperature in detail. For antiferromagnetic case the $T$-dependence of the discords exhibits exponential damping behavior with increasing $T$ like thermal entanglement, but  it does not
reach to exact zero.
For ferromagnetic case, however, thermal discord becomes exact zero at the point  ${\bm D} = {\bm D_*}$  and $T_c=0$. The behavior of thermal discord
at $T = T_c$ for arbitrary $|{\bm D}|$ is also examined in detail. 
In section V a brief conclusion is given. In appendix A the thermal discord for $\rho_Y (T)$ are explicitly computed. In appendix B we discuss the 
critical behavior of concurrence when $D_x \neq 0$ and $D_y \neq 0$ for completeness.

\section{Thermal Density Matrix}
The Hamiltonian for two-spin anisotropic Heisenberg $X$ $Y$ $Z$ chain with DM interaction is given by\cite{byrnes17}
\begin{eqnarray}
\label{hamil-total}
H = J_x \sigma_1^x \otimes \sigma_2^x + J_y \sigma_1^y \otimes \sigma_2^y + J_z \sigma_1^z \otimes \sigma_2^z 
+ {\bm D} \cdot \left( {\bm \sigma}_1 \times {\bm \sigma}_2 \right).
\end{eqnarray}
In Ref. \cite{guof1} and Ref. \cite{cao08,langari09} the entanglement of this model with $J_x = J_y = J_z$ and $J_x = J_y \neq J_z$ was discussed respectively. 
The quantum phase transition with an applied magnetic field was studied in Ref. \cite{li08}. Also the thermal entanglement of three-qubit ground state was 
discussed\cite{jafari11,marko-5}. Quantum discord\cite{discord1,discord2}, another measure of quantum correlation, of this model was discussed in 
Ref. \cite{zhi10,zidan1} when $J_x = J_y$ and in Ref. \cite{zou11} when $J_z = 0$. Recently, the quantum-memory-assisted entropic
uncertainties\cite{renes09,berta10} were examined in this Heisenberg model\cite{zhang18}.

\subsection{$D_x = D_y = 0$ case}
In this case the matrix representation of the Hamiltonian in the computational basis $\{ \ket{00}, \ket{01}, \ket{10}, \ket{11} \}$ becomes
\begin{eqnarray}
\label{hamil-Z}
H_Z = \left(           \begin{array}{cccc}
                 J_z  &  0  &  0  &   J_x - J_y                                \\
                 0  & -J_z  &  J_x + J_y + 2 i D_z  &  0                    \\
                 0  &  J_x + J_y - 2 i D_z  &  -J_z  &  0                   \\
                 J_x - J_y  &  0  &  0  &  J_z
                            \end{array}                                          \right).
\end{eqnarray}
The eigenvalues and corresponding eigenvectors of $H_Z$ are summarized in Table I. In this Table $\xi$ and $\theta$ are given by
\begin{equation}
\label{table1}
\xi = \sqrt{4 D_z^2 + (J_x + J_y)^2}   \hspace{1.0cm}  \theta = \tan^{-1} \left(-\frac{2 D_z}{J_x + J_y} \right).
\end{equation}

\begin{center}
\begin{tabular}{c|c} \hline \hline
eigenvalues of $H_Z$ & corresponding eigenvectors                                     \\  \hline \hline
$E_{1,z} = J_x - J_y + J_z$ &  $\ket{z_1} = \frac{1}{\sqrt{2}} \left(\ket{00} + \ket{11} \right)$                    \\    
$E_{2,z} = -J_x + J_y + J_z$ & $\ket{z_2} = \frac{1}{\sqrt{2}} \left(\ket{00} - \ket{11} \right)$                    \\  
$E_{3,z} = - J_z + \xi$  &  $\ket{z_3} = \frac{1}{\sqrt{2}} \left( \ket{01} + e^{i \theta} \ket{10} \right)$                \\
$E_{4,z} = - J_z - \xi$  &  $\ket{z_4} = \frac{1}{\sqrt{2}} \left( \ket{01} - e^{i \theta} \ket{10} \right)$       \\ \hline
\end{tabular}

\vspace{0.3cm}
Table I: Eigenvalues and eigenvectors of $H_Z$ 
\end{center}
\vspace{0.5cm}
Thus, the spectral decomposition of $H_Z$ can be written as 
\begin{equation}
\label{optimal-z}
H_Z = \sum_{i=1}^4 E_{i, z} \ket{z_i} \bra{z_i}.
\end{equation}

The partition function of this system is given by 
\begin{equation}
\label{partition-z}
{\cal Z}_Z \equiv \mbox{Tr} \left[ e^{-\beta H_Z} \right] = 2 e^{-\beta J_z} \cosh\left[\beta (J_x - J_y) \right] 
+ 2 e^{\beta J_z} \cosh (\beta \xi)
\end{equation}
where $\beta = 1 / k_B T$, where $k_B$ and $T$ are Boltzmann constant and external temperature. Throughout this paper we use $k_B = 1$ for convenience.
Then the thermal density matrix in this case becomes
\begin{eqnarray}
\label{density-z}
\rho_Z (T) \equiv \frac{1}{{\cal Z}_Z} e^{-\beta H_Z} 
= \left(             \begin{array}{cccc}
             r  &  0  &  0  &  s                          \\
             0  &  u  &  v  &  0                          \\
             0  &  v^*  &  u  &  0                       \\
             s  &  0  &  0  &  r
                        \end{array}                                  \right)  
\end{eqnarray}
where
\begin{eqnarray}
\label{density-z-boso}
&&r = \frac{1}{{\cal Z}_Z} e^{-\beta J_z} \cosh \left[ \beta (J_x - J_y) \right]               \hspace{1.0cm}
s = -\frac{1}{{\cal Z}_Z} e^{-\beta J_z} \sinh \left[ \beta (J_x - J_y) \right]                           \\    \nonumber
&& u =  \frac{1}{{\cal Z}_Z} e^{\beta J_z} \cosh (\beta \xi)                                          \hspace{1.0cm}
v = - \frac{1}{{\cal Z}_Z} \frac{J_x + J_y + 2 i D_z}{\xi} e^{\beta J_z} \sinh (\beta \xi).
\end{eqnarray}
It is worthwhile noting $r + u = 1/2$, which guarantees $\mbox{Tr} [\rho_Z (T)] = 1$. 

\subsection{$D_x = D_z = 0$ case}
In this case the matrix representation of the Hamiltonian in the computational basis becomes
\begin{eqnarray}
\label{hamil-Y}
H_Y = \left(           \begin{array}{cccc}
                 J_z  &  D_y  &  -D_y  &   J_x - J_y                                \\
                 D_y  & -J_z  &  J_x + J_y   &  D_y                    \\
                 -D_y  &  J_x + J_y &  -J_z  &  -D_y                   \\
                 J_x - J_y  &  D_y  &  -D_y  &  J_z
                            \end{array}                                          \right).
\end{eqnarray}
The eigenvalues and corresponding eigenvectors of $H_Y$ are summarized in Table II. In this Table $\eta$, $\phi_1$ and $\phi_2$ are given by
\begin{equation}
\label{table2}
\eta = \sqrt{4 D_y^2 + (J_x + J_z)^2}   \hspace{.5cm}  \phi_1 = \tan^{-1} \left[-\frac{2 D_y}{\eta - (J_x + J_z)} \right] 
\hspace{.5cm}  \phi_2 = \tan^{-1} \left[\frac{2 D_y}{\eta + (J_x + J_z)} \right].
\end{equation}

\begin{center}
\begin{tabular}{c|c} \hline \hline
eigenvalues of $H_Y$ & corresponding eigenvectors                                     \\  \hline \hline
$E_{1,y} = J_y + (J_x - J_z)$ &  $\ket{y_1} = \frac{1}{\sqrt{2}} \left(\ket{01} + \ket{10} \right)$                    \\    
$E_{2,y} = J_y - (J_x - J_z) $ & $\ket{y_2} = \frac{1}{\sqrt{2}} \left(\ket{00} - \ket{11} \right)$                    \\  
$E_{3,y} = - J_y + \eta$  &  $\ket{y_3} = \frac{1}{\sqrt{2}} \left[ \sin \phi_1 (\ket{00} + \ket{11}) - \cos \phi_1 (\ket{01} - \ket{10}) \right]$                       
                                                                                                                                                                      \\
$E_{4,y} = - J_y - \eta$  &  $\ket{y_4} =  \frac{1}{\sqrt{2}} \left[ \sin \phi_2 (\ket{00} + \ket{11}) - \cos \phi_2 (\ket{01} - \ket{10}) \right]$                           
                                                                                                                                                                    \\ \hline
\end{tabular}

\vspace{0.3cm}
Table II: Eigenvalues and eigenvectors of $H_Y$ 
\end{center}
\vspace{0.5cm}
The fact $\cos (\phi_1 - \phi_2) = 0$ guarantees the orthonormal condition of $\ket{y_j}$, i.e., $\bra{y_i}y_j\rangle = \delta_{ij}$. 

Thus, the spectral decomposition of $H_Y$ is $H_Y = \sum_{i=1}^4 E_{i,y} \ket{y_i} \bra{y_i}$ and the partition function is 
\begin{equation}
\label{partition-y}
{\cal Z}_Y \equiv \mbox{Tr} \left[ e^{-\beta H_Y} \right] = 2 e^{-\beta J_y} \cosh \left[\beta (J_x - J_z) \right] 
+ 2 e^{\beta J_y} \cosh (\beta \eta).
\end{equation}
The thermal density matrix in this system can be written in a form
\begin{eqnarray}
\label{density-y}
\rho_Y (T) = \frac{1}{{\cal Z}_Y} e^{-\beta H_Y} =   \left(                          \begin{array}{cccc}
                                                                                                                 r_1  &  -q  &  q  &  r_2                        \\
                                                                                                                 -q  &  u_1  &  u_2  &  -q                     \\
                                                                                                                 q  &  u_2  &  u_1  &  q                        \\
                                                                                                                 r_2  &  -q  &  q  &  r_1
                                                                                                                   \end{array}                              \right)
\end{eqnarray}
where
\begin{eqnarray}
\label{density-y-boso}
&&r_1 = \frac{1}{2 {\cal Z}_Y} \left[ e^{-\beta E_{2,y}} + \sin^2 \phi_1 e^{-\beta E_{3,y}} + \sin^2 \phi_2 e^{-\beta E_{4,y}} \right]
                                                                                                                                                                                                \\    \nonumber
&&r_2 = \frac{1}{2 {\cal Z}_Y} \left[- e^{-\beta E_{2,y}} + \sin^2 \phi_1 e^{-\beta E_{3,y}} + \sin^2 \phi_2 e^{-\beta E_{4,y}} \right]
                                                                                                                                                                                                \\    \nonumber
&&u_1 = \frac{1}{2 {\cal Z}_Y} \left[ e^{-\beta E_{1,y}} + \cos^2 \phi_1 e^{-\beta E_{3,y}} + \cos^2 \phi_2 e^{-\beta E_{4,y}} \right]
                                                                                                                                                                                                \\    \nonumber
&&u_2 = \frac{1}{2 {\cal Z}_Y} \left[ e^{-\beta E_{1,y}} - \cos^2 \phi_1 e^{-\beta E_{3,y}} - \cos^2 \phi_2 e^{-\beta E_{4,y}} \right]
                                                                                                                                                                                                \\    \nonumber
&&q = \frac{1}{2 {\cal Z}_Y} \left[ \sin \phi_1 \cos \phi_1 e^{-\beta E_{3, y}} + \sin \phi_2 \cos \phi_2 e^{-\beta E_{4, y}} \right].
\end{eqnarray}
Since $r_1 + u_1 = 1/2$, it is easy to show $\mbox{Tr} \left[ \rho_Y (T) \right] = 1$. Since $D_y = D_z = 0$ case is similar to 
$D_x = D_z = 0$ case, we do not explore this case in this paper.

\section{Thermal Entanglement}
In this section we examine the temperature dependence of entanglement for the thermal density matrices $\rho_Z (T)$ and $\rho_Y (T)$ by making use of 
concurrence\cite{form2,form3}. Thermal entanglement of $X$ $Y$ $Z$ model was considered in Ref. \cite{rigolin04} when there is no DM interaction and 
in Ref. \cite{cao-5} when one component of DM-interaction is nonzero. Thus, our calculation of this section overlaps with that of Ref. \cite{cao-5}. 
However we focus on the critical temperature $T_c$ in detail because our main motivation is to examine the quantum discord around $T \sim T_c$
Following the procedure presented in Ref. \cite{form3} one can compute the concurrence 
${\cal C} (\rho)$ for a quantum state $\rho$ by a simple formula
\begin{equation}
\label{two4}
{\cal C} (\rho) = \mbox{max} \left(\lambda_1 - \lambda_2 - \lambda_3 - \lambda_4, 0\right),
\end{equation}
where the $\lambda_i$'s are eigenvalues, in decreasing order, of the Hermitian matrix
$\sqrt{\sqrt{\rho} (\sigma_y \otimes \sigma_y) \rho^* (\sigma_y \otimes \sigma_y) \sqrt{\rho}}$.

Following this procedure it is easy to show that the concurrence of $\rho_Z (T)$ is 
\begin{equation}
\label{concur-z1}
{\cal C} \left( \rho_Z \right) = \max \left[ |\xi_1 - \xi_3| - \xi_2 - \xi_4, 0 \right]
\end{equation}
where
\begin{eqnarray}
\label{concur-z2}
&&\xi_1 = \frac{1}{{\cal Z}_Z} e^{\beta [|J_x - J_y| - J_z]}    \hspace{2.0cm} \xi_2 = \frac{1}{{\cal Z}_Z} e^{-\beta [|J_x - J_y| + J_z]}
                                                                                                                                                                                      \\    \nonumber
&&\xi_3 =  \frac{1}{{\cal Z}_Z} e^{\beta (J_z + \xi)}   \hspace{2.0cm}  \xi_4 =  \frac{1}{{\cal Z}_Z} e^{\beta (J_z - \xi)}.
\end{eqnarray}

\begin{figure}[ht!]
\begin{center}
\includegraphics[height=5.0cm]{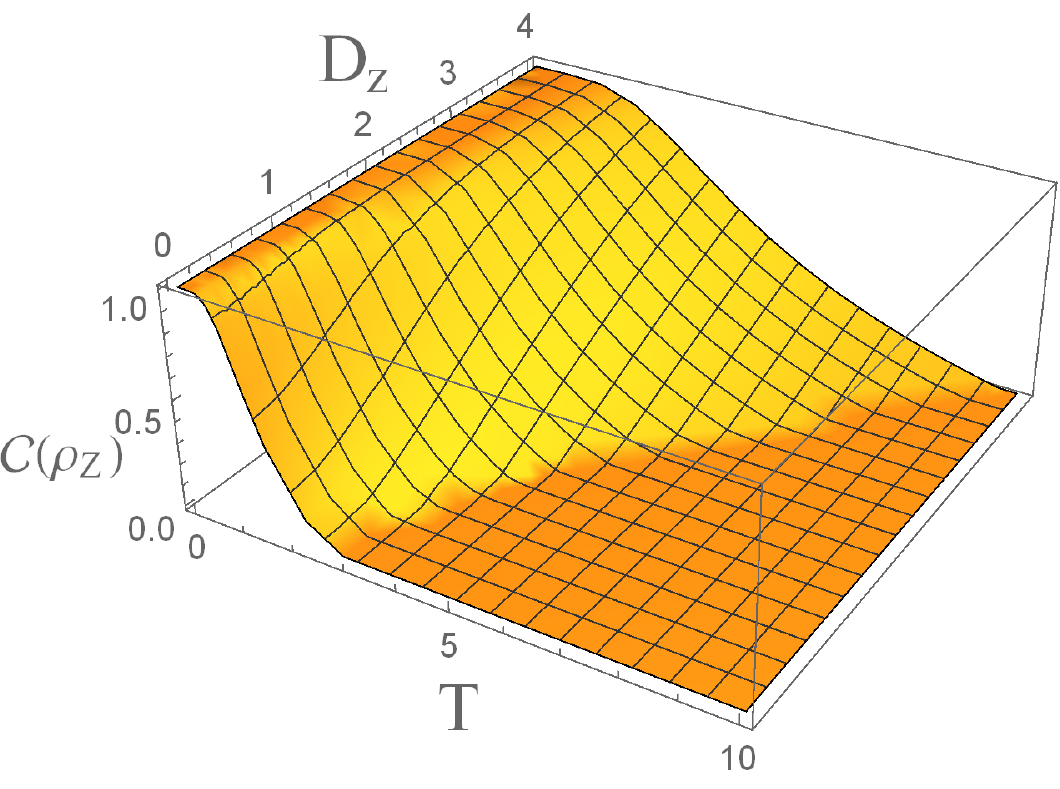} 
\includegraphics[height=5.0cm]{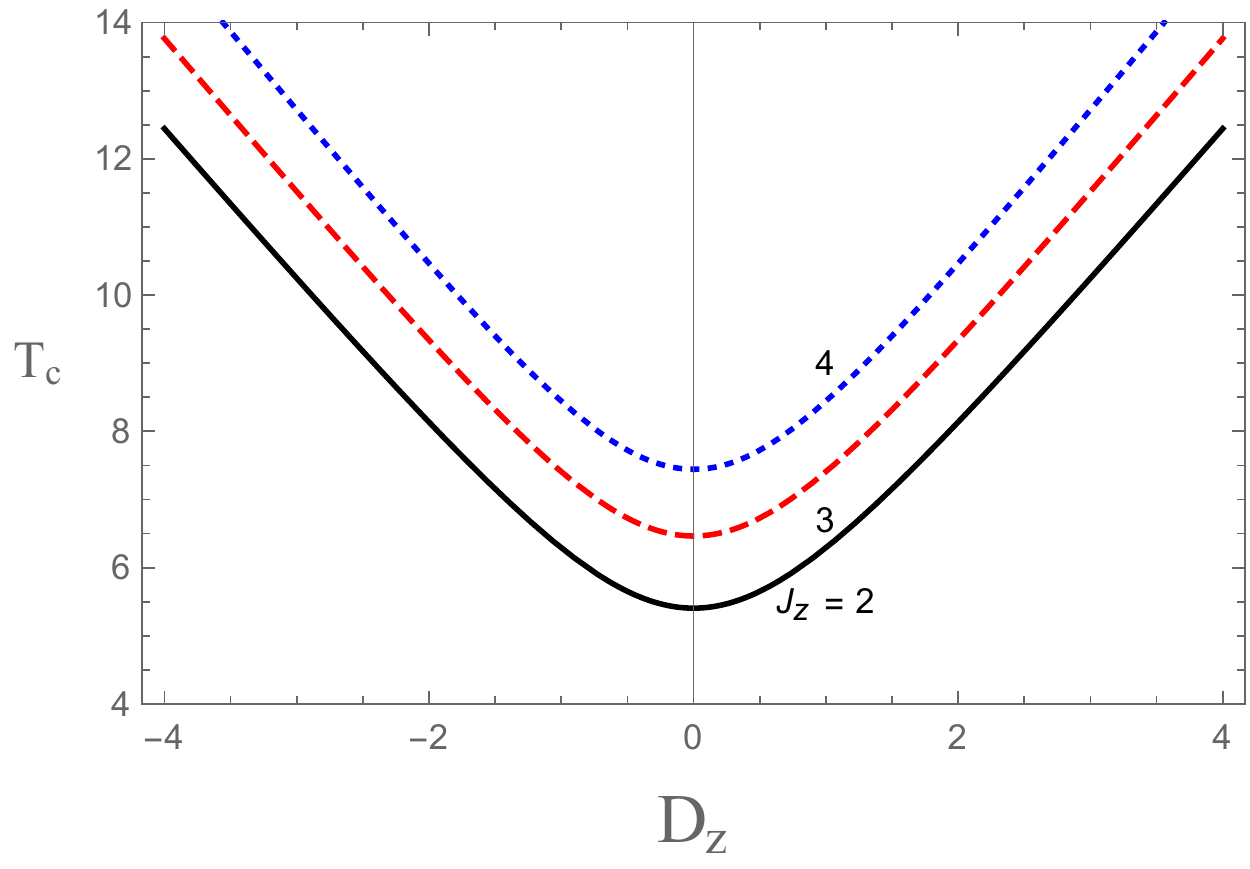}

\caption[fig1]{(Color online) (a) The $T$- and $D_z$-dependence of concurrence ${\cal C} (\rho_Z)$ 
in $X$ $X$ $Z$ model with DM interaction when $J=1$ and $J_z = 0.2$. 
This figure shows that this model exhibits a quantum phase transition with the critical temperature $T_c$, where the entanglement completely vanishes 
at $T \geq T_c$. (b) The $D_z$-dependence of $T_c$  when $J_z =2$ (black line), $J_z = 3$ (red dashed line), and $J_z = 4$ (blue dotted line) with choosing 
$J_x = 1$ and $J_y = 1.5$. This figure shows that the critical temperature increases very rapidly 
with increasing $|D_z|$.
 }
\end{center}
\end{figure}

Before proceeding further, let us consider $X$ $X$ $Z$ model ($J_x = J_y = J$) with DM interaction. In this model the concurrence becomes
\begin{equation}
\label{concur-xxz-z}
{\cal C} \left( \rho_Z \right) = \frac{e^{\beta J_z}}{{\cal Z}_Z} \max \left[ |e^{2 \beta w} - e^{-2 \beta J_z}| - e^{-2 \beta w} - e^{-2 \beta J_z}, 0
\right]
\end{equation}
where $w = \sqrt{J^2 + D_z^2}$. This is plotted in Fig. 1(a) as a function of $T$ and $D_z$ with choosing $J=1$ and $J_z = 0.2$. As this figure shows, this 
model exhibits a quantum phase transition with the critical temperature $T_c$. This means that the entanglement completely vanishes at $T \geq T_c$. 
From Eq. (\ref{concur-xxz-z}) one can show that  the critical temperature satisfies
\begin{equation}
\label{critical-1}
e^{2 J_z / T_c} \sinh \left( \frac{2 w}{T_c} \right) = 1.
\end{equation}

\begin{figure}[ht!]
\begin{center}
\includegraphics[height=5.0cm]{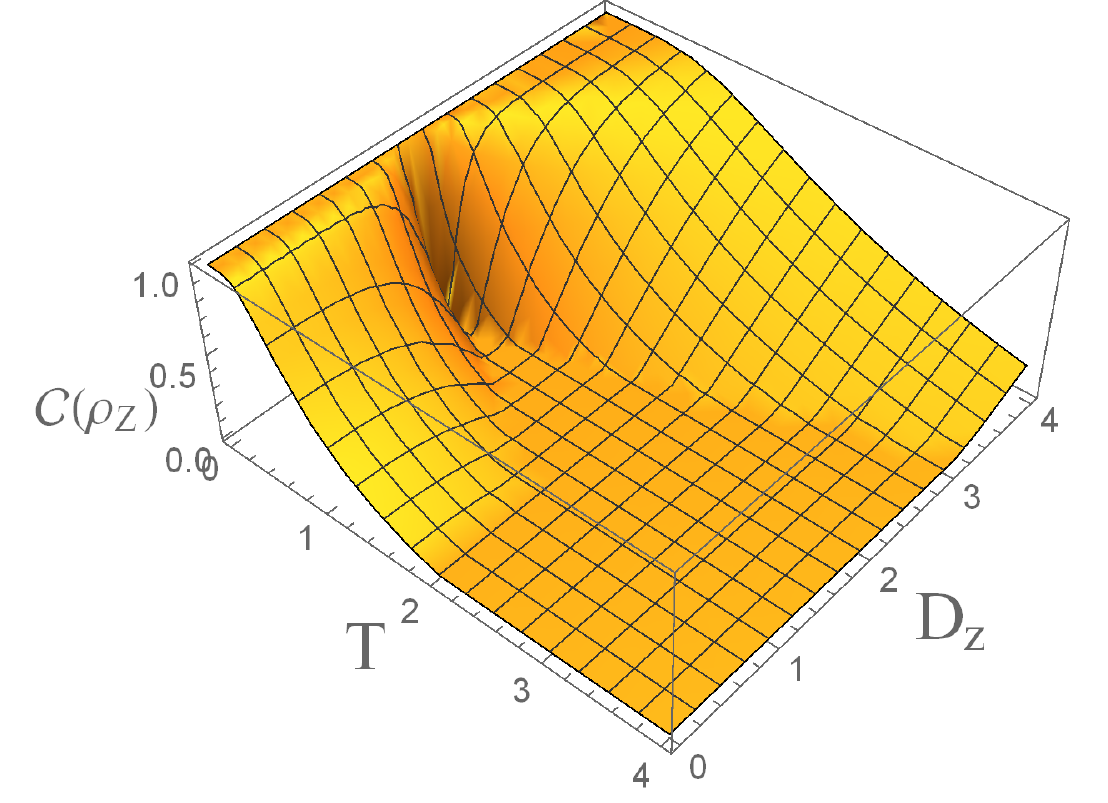} 
\includegraphics[height=5.0cm]{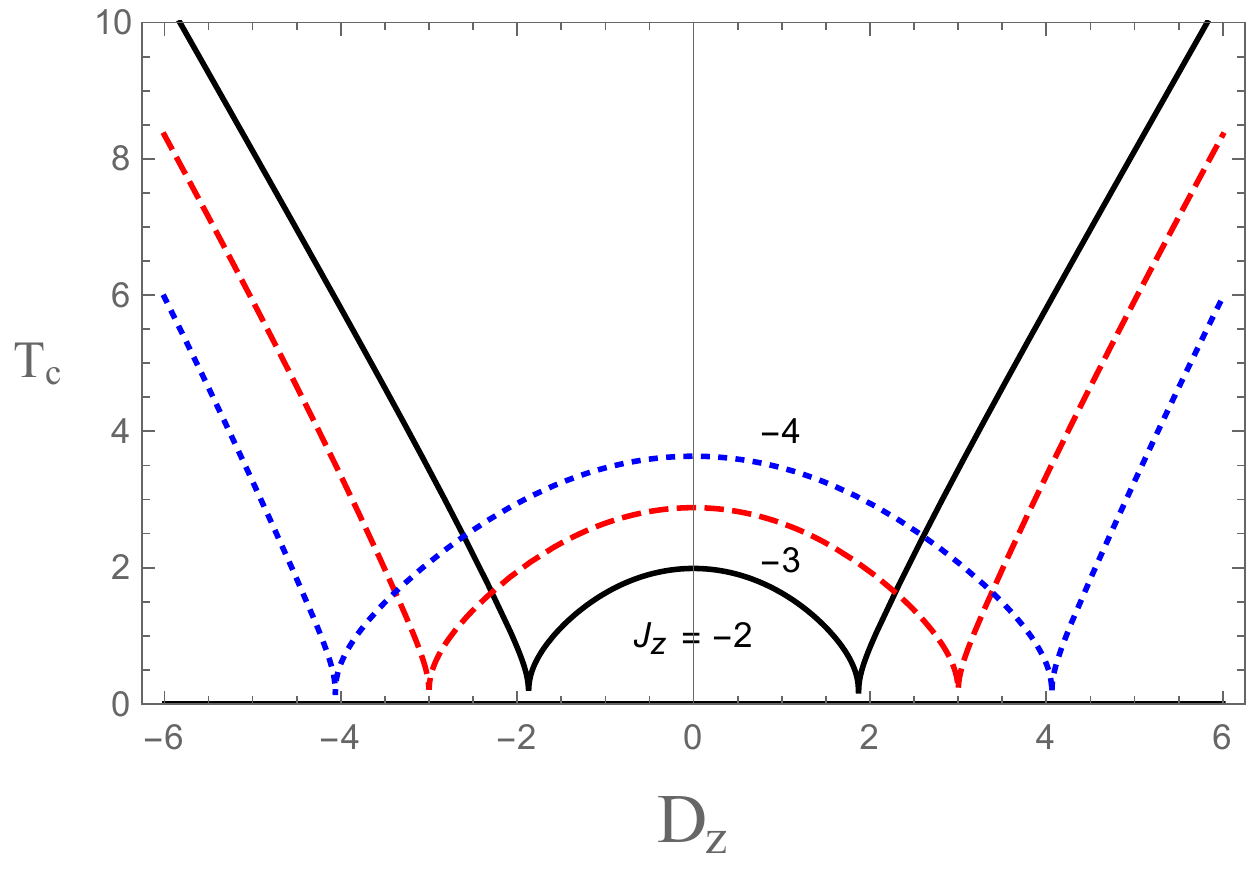}

\caption[fig2]{(Color online) (a) The $T$- and $D_z$-dependence of concurrence ${\cal C} (\rho_Z)$  in $X$ $Y$ $Z$ model with DM interaction when $J_x = -1$, 
$J_y = -1.5$, and $J_z = -2$.  As this figure shows, the concurrence exhibits different behavior in 
$|D_z| \leq \sqrt{7/2} = 1.87$ and other regions. This is because of the fact that the critical temperature is determined by different equations in these 
regions (see Eq. (\ref{critical-2})).  (b)  The $D_z$-dependence of the critical temperature $T_c$  when  $J_z = -2$ (black line), $J_z = -3$ (red dashed line),
and $J_z = -4$ (blue dotted line) with choosing $J_x = -1$ and $J_y = -1.5$. As this figure shows, $T_c$ increases monotonically with increasing 
$|D_z|$ in the region $|D_z| \geq \sqrt{7/2}$ as $X$ $X$ $Z$ model. However, it behaves differently in the region $|D_z| < \sqrt{7/2}$ as Fig. 2(a) exhibits.}
\end{center}
\end{figure}

Now, let us go back to the $X$ $Y$ $Z$ model with DM interaction. In this model the critical temperature $T_c$ satisfies
\begin{eqnarray}
\label{critical-2}
&& e^{2 J_z / T_c} \frac{\sinh  (\xi / T_c )}{\cosh (|J_x - J_y| / T_c)} = 1   \hspace{1.0cm} \mbox{when} \hspace{.3cm}  2 J_z + \xi \geq |J_x - J_y|   \\   \nonumber
&& e^{-2 J_z / T_c} \frac{\sinh  (|J_x - J_y| / T_c )}{\cosh (\xi / T_c)} = 1  \hspace{1.0cm} \mbox{when}  \hspace{.3cm} 2 J_z + \xi <  |J_x - J_y|.
\end{eqnarray}
When $J_x = J_y$, first equation reduces to Eq. (\ref{critical-1}) and second equation does not yield any solution. 

For antiferromagnetic ($J_{\alpha} > 0$)
case the second equation does not play any role because the condition $2 J_z + \xi < |J_x - J_y|$ cannot be satisfied. Thus, the behavior of 
the concurrence ${\cal C} (\rho_Z)$ is similar to that of $X$ $X$ $Z$ model. The $D_z$-dependence of $T_c$ is plotted in Fig. 1(b) when $J_x = 1$
and $J_y = 1.5$ with varying $J_z = 2$ (black line), $3$ (red dashed line), and $4$ (blue dotted line). Fig. 1(b) shows that $T_c$ increases very rapidly
with increasing $|D_z|$. With fixed $D_z$, $T_c$ increases with increasing $J_z$. 

For ferromagnetic  ($J_{\alpha} < 0$) case, however, the second equation of Eq. (\ref{critical-2})
provides significant solutions, which result in different behavior of  ${\cal C} (\rho_Z)$. The boundary of two region in Eq. (\ref{critical-2}) is  $D_z = D_{z,*}$, where 
\begin{equation}
\label{dzstar}
D_{z,*} = \sqrt{(J_z - J_>) (J_z + J_<)}
\end{equation}
with $J_> = \max(J_x, J_y)$ and $J_< = \min(J_x, J_y)$. At this point $T_c$ becomes exactly zero because $\xi_1 = \xi_3$ at this point. 
In Fig. 2(a) we plot  ${\cal C} (\rho_Z)$ as a function of 
$T$ and $D_z$ with choosing $J_x = -1$, $J_y = -1.5$, and $J_z = -2$. As this figure shows, the concurrence exhibits different behavior in 
$0 \leq D_z \leq \sqrt{7/2} = 1.87$ and other regions. This is because of the fact that the second equation of Eq. (\ref{critical-2}) generates the critical 
temperature in the region $0 \leq D_z \leq \sqrt{7/2}$ while the first equation generates it in other region. In order to show more precisely we plot the 
$D_z$-dependence of $T_c$ in Fig. 2 (b) when  $J_z = -2$ (black line), $ -3$ (red dashed line), and $-4$ (blue dotted line) with fixing $J_x = -1$ and 
$J_y = -1.5$.  As this figure shows, $T_c$ increases monotonically with increasing 
$|D_z|$ in the region $|D_z| \geq D_{z,*} $ as $X$ $X$ $Z$ model. With fixed $D_z$, in this region, $T_c$ decreases with increasing $|J_z|$.
However, it behaves differently in the region $|D_z| < D_{z,*}$. In this region $T_c$ decreases with increasing $|D_z|$. With fixed $D_z$, in this region, 
$T_c$ increases with increasing $|J_z|$.  At $D_z = D_{z,*}$ this figure confirms $T_c = 0$. 

\begin{figure}[ht!]
\begin{center}
\includegraphics[height=5.0cm]{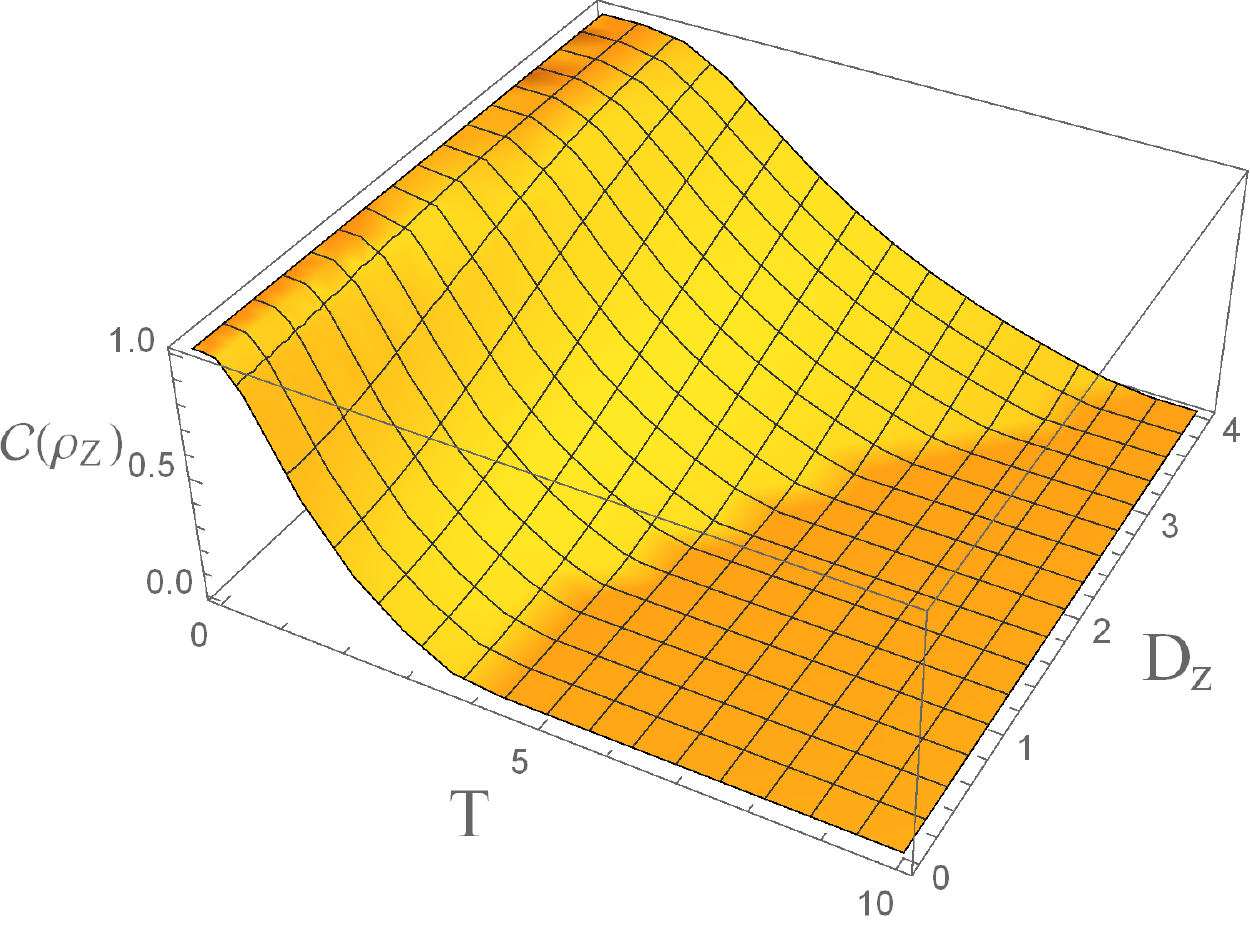} 
\includegraphics[height=5.0cm]{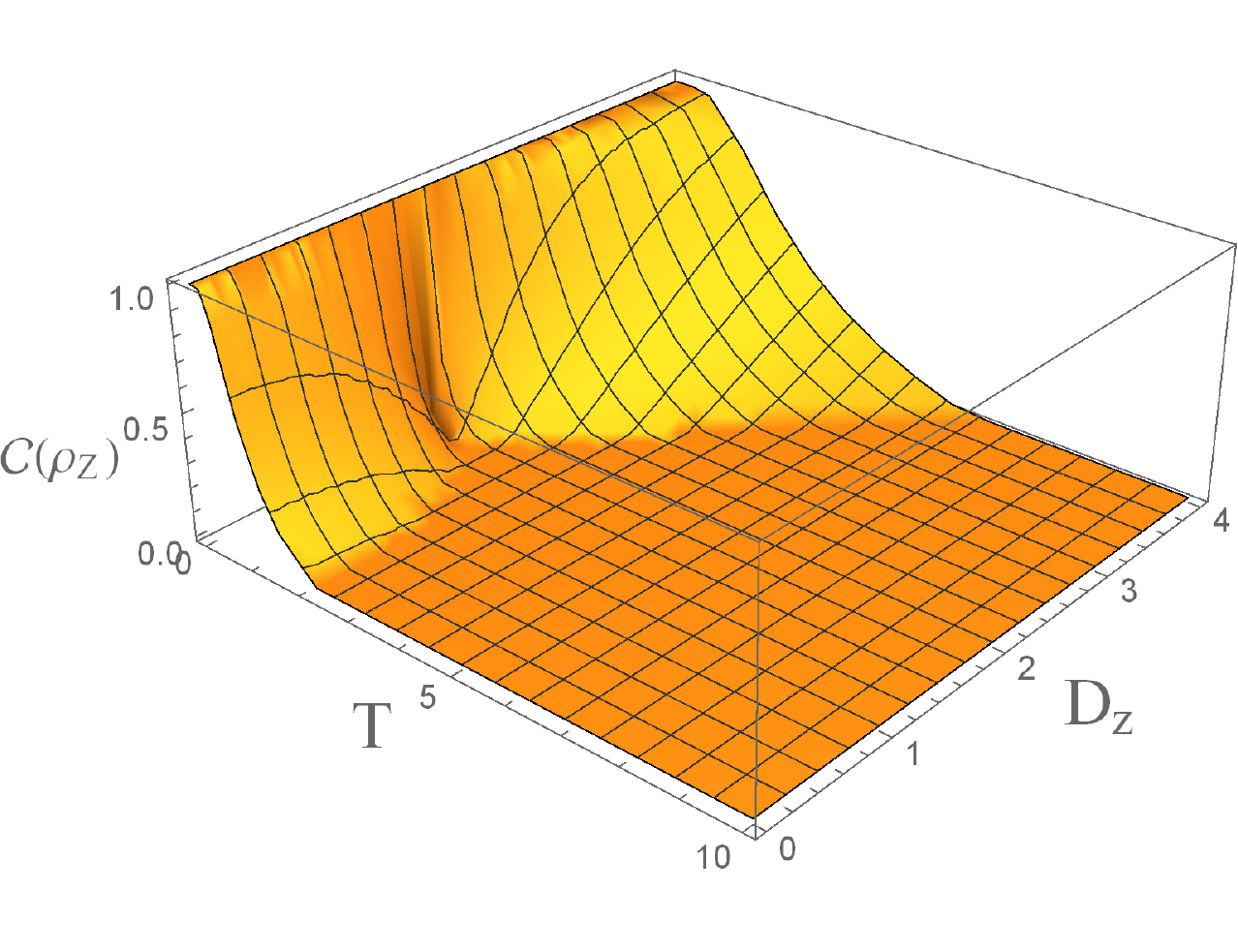}

\caption[fig]{(Color online)  The $T$- and $D_z$- dependence of ${\cal C} (\rho_Z)$ when  $J_x = 3$ and  $J_y = 2$ with $J_z = -1$ (Fig. 3 (a)) and $J_z = -2.5$ (Fig. 3(b)). As expected Fig. 3(a) and Fig. 3(b) exhibit similar behavior to Fig. 1(a) and Fig. 2(a) respectively. This is due to the existence or nonexistence of $D_{z,*}$. }
\end{center}
\end{figure}

For neither ferromagnetic nor antiferromagnetic case the behavior of the concurrence is determined whether the second equation of Eq. (\ref{critical-2}) plays a role or not. For example, let us consider $J_z > 0$ case. If, in this case, $D_z^2 + J_x J_y \geq 0$, the second equation of Eq. (\ref{critical-2}) 
does not play any role. Thus, in this case the concurrence is similar to Fig. 1 (a). If $D_z^2 + J_x J_y <0$, the condition for the existence of $D_{z,*}$ is 
$0 < J_z < \min (-J_<, J_>)$. Thus, both  $D_z^2 + J_x J_y <0$ and $0 < J_z < \min (-J_<, J_>)$ hold, the concurrence of $\rho_Z$ is similar to Fig. 2 (a).
For negative $J_z$ case one can derive $J_z < \min (-J_<, 0)$ for $J_> > - J_<$ and $J_z < \min (J_>, 0)$ for $-J_< > J_>$ for the existence of $D_{z,*}$.
In Fig. 3 we plot $D_z$- and $T$-dependence of ${\cal C} (\rho_Z)$ when $J_x = 3$ and  $J_y = 2$ with $J_z = -1$ (Fig. 3 (a)) and $J_z = -2.5$ (Fig. 3(b)). 
As expected Fig. 3(a) and Fig. 3(b) exhibit similar behavior to Fig. 1(a) and Fig. 2(a) respectively.

\begin{figure}[ht!]
\begin{center}
\includegraphics[height=5.0cm]{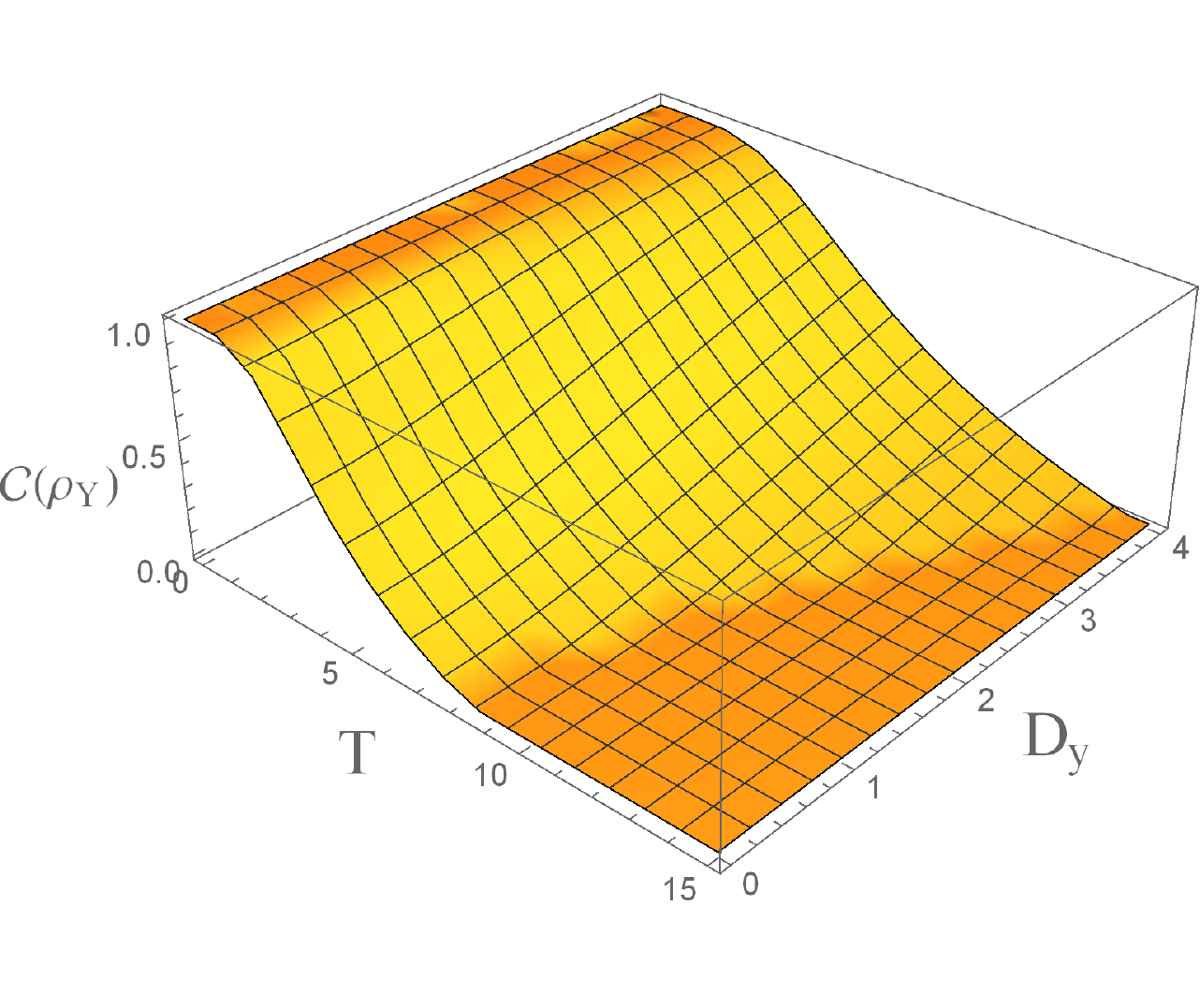} 
\includegraphics[height=5.0cm]{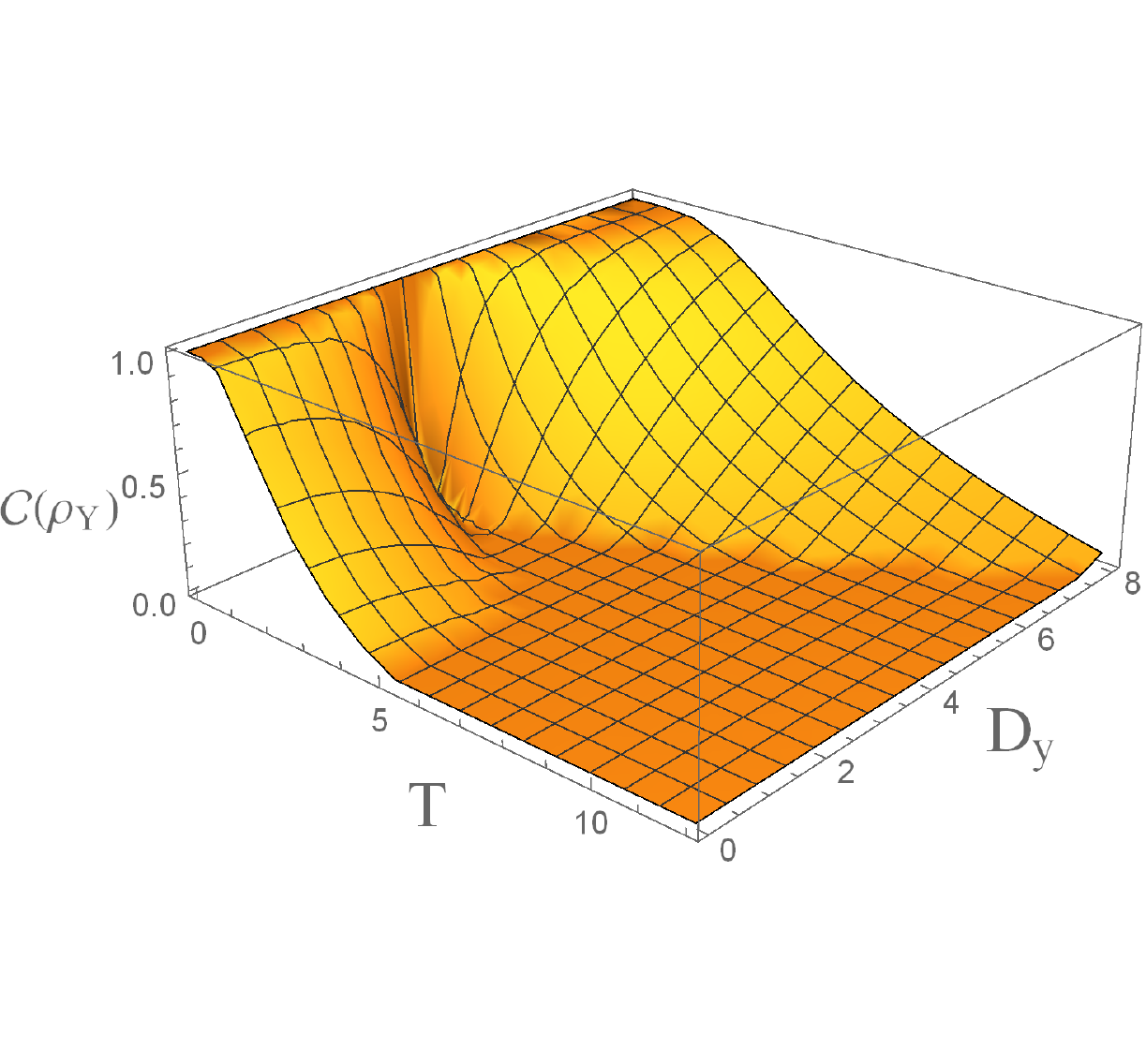}
\includegraphics[height=5.0cm]{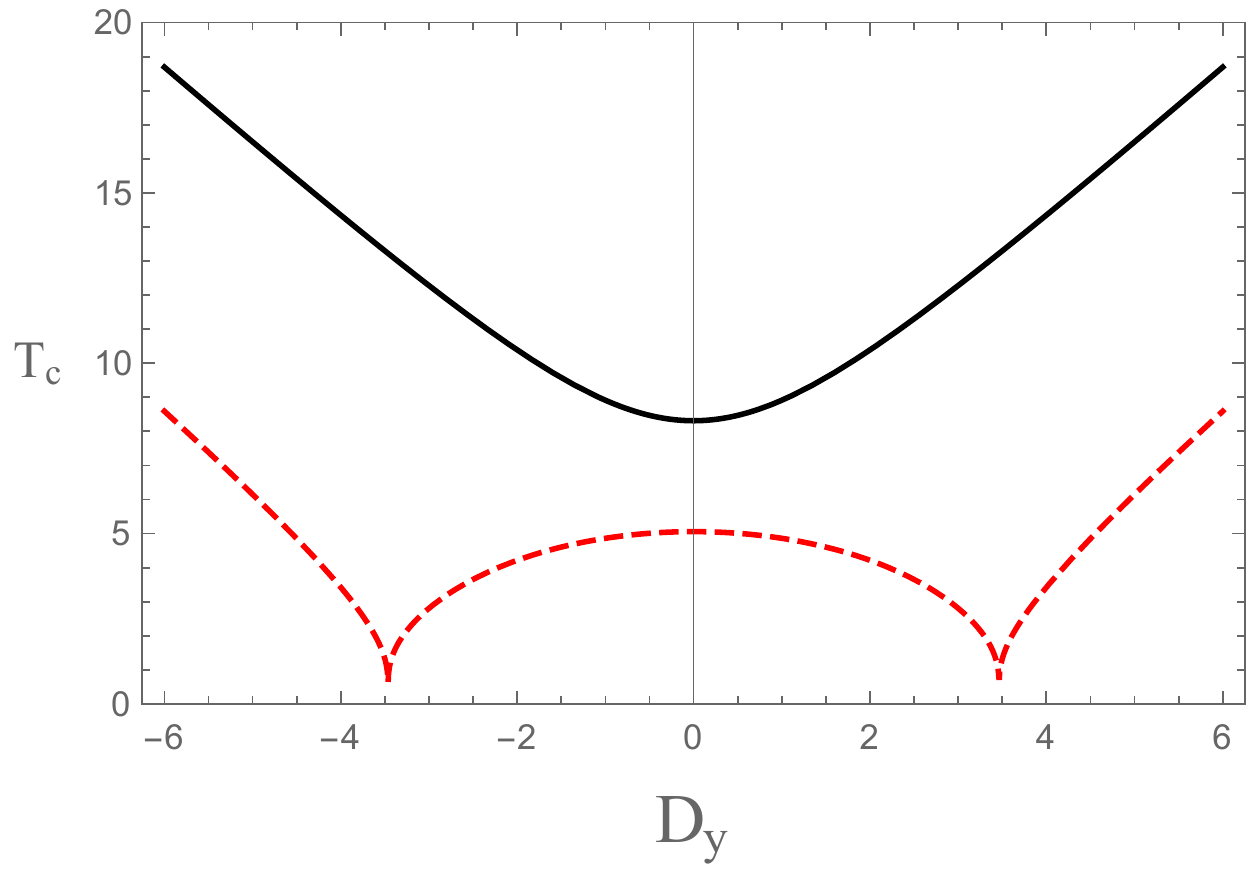}

\caption[fig4]{(Color online) (a) The $T$- and $D_y$- dependence of ${\cal C} (\rho_Y)$ when $J_x = J_y = 3$ and $J_z = 1$. This behaves very similar to 
Fig. 1(a). This is because of the fact that the critical temperature is determined by single equation. (b) The $T$- and $D_y$- dependence of ${\cal C} (\rho_Y)$
when $J_x = J_y = -3$ and $J_z = -1$. This behaves very similar to Fig. 2(a). This is because of the fact that the critical temperature is determined by two
different equations. (see Eq. (\ref{critical-3})) (c)  The $D_y$-dependence of $T_c$  for antiferromagnetic ($J= 3, J_z = 1$) and ferromagnetic ($J= -3, J_z = -1$) cases as black and red dashed lines, respectively.  }
\end{center}
\end{figure}

For $\rho_Y (T)$ the concurrence becomes 
\begin{equation}
\label{concur-y1}
{\cal C} \left( \rho_Y \right) = \max \left[ |\eta_1 - \eta_3| - \eta_2 - \eta_4, 0 \right]
\end{equation}
where
\begin{eqnarray}
\label{concur-y2}
&&\eta_1 = \frac{1}{{\cal Z}_Y} e^{\beta [|J_x - J_z| - J_y]}    \hspace{2.0cm} \eta_2 = \frac{1}{{\cal Z}_Y} e^{-\beta [|J_x - J_z| + J_z]}
                                                                                                                                                                                      \\    \nonumber
&&\eta_3 =  \frac{1}{{\cal Z}_Y} e^{\beta (J_y + \eta)}   \hspace{2.0cm}  \eta_4 =  \frac{1}{{\cal Z}_Y} e^{\beta (J_y - \eta)}.
\end{eqnarray}
Unlike $\rho_Z (T)$, even in the $X$ $X$ $Z$ model with DM interaction the critical temperature $T_c$ is determined by two different 
equations as follows
\begin{eqnarray}
\label{critical-3}
&& e^{2 J / T_c} \frac{\sinh  (\eta / T_c )}{\cosh (|J - J_z| / T_c)} = 1   \hspace{1.0cm} \mbox{when} \hspace{.3cm}  2 J + \eta \geq |J - J_z|                                
                                                                                                                                                                              \\   \nonumber
&& e^{-2 J / T_c} \frac{\sinh  (|J - J_z| / T_c )}{\cosh (\eta / T_c)} = 1  \hspace{1.0cm} \mbox{when}  \hspace{.3cm} 2 J + \eta <  |J - J_z|.
\end{eqnarray}
For the antiferromagnetic case ($J, J_z > 0$) the critical temperature is determined by only first equation of Eq. (\ref{critical-3}) because 
$2 J + \eta \geq |J - J_z|$ in this case. Thus, the concurrence exhibits similar behavior with that of Fig. 1(a). This is confirmed in Fig. 4(a), where 
${\cal C} (\rho_Y)$ is plotted as a function of $T$ and $D_y$ with choosing $J= 3$ and $J_z = 1$. For the ferromagnetic ($J, J_z <0$) case, however, 
both equations of Eq. (\ref{critical-3}) play significant role for determining the critical temperature. This is also confirmed in Fig. 4(b), where 
${\cal C} (\rho_Y)$ is plotted as a function of $T$ and $D_y$ with choosing $J= -3$ and $J_z = -1$. In this case the second equation of Eq. (\ref{critical-3})
determines $T_c$ in the region $|D_y| < 2 \sqrt{3} = 3.46$ while $T_c$ in other region is determined by the first equation of Eq. (\ref{critical-3}). 
In Fig. 4(c) the $D_y$-dependence of $T_c$ is plotted for antiferromagnetic ($J= 3, J_z = 1$) and ferromagnetic ($J= -3, J_z = -1$) cases as 
black and red dashed lines, respectively. 
As expected, the critical temperature for ferromagnetic case behaves differently from that for antiferromagnetic case.

\begin{figure}[ht!]
\begin{center}
\includegraphics[height=5.0cm]{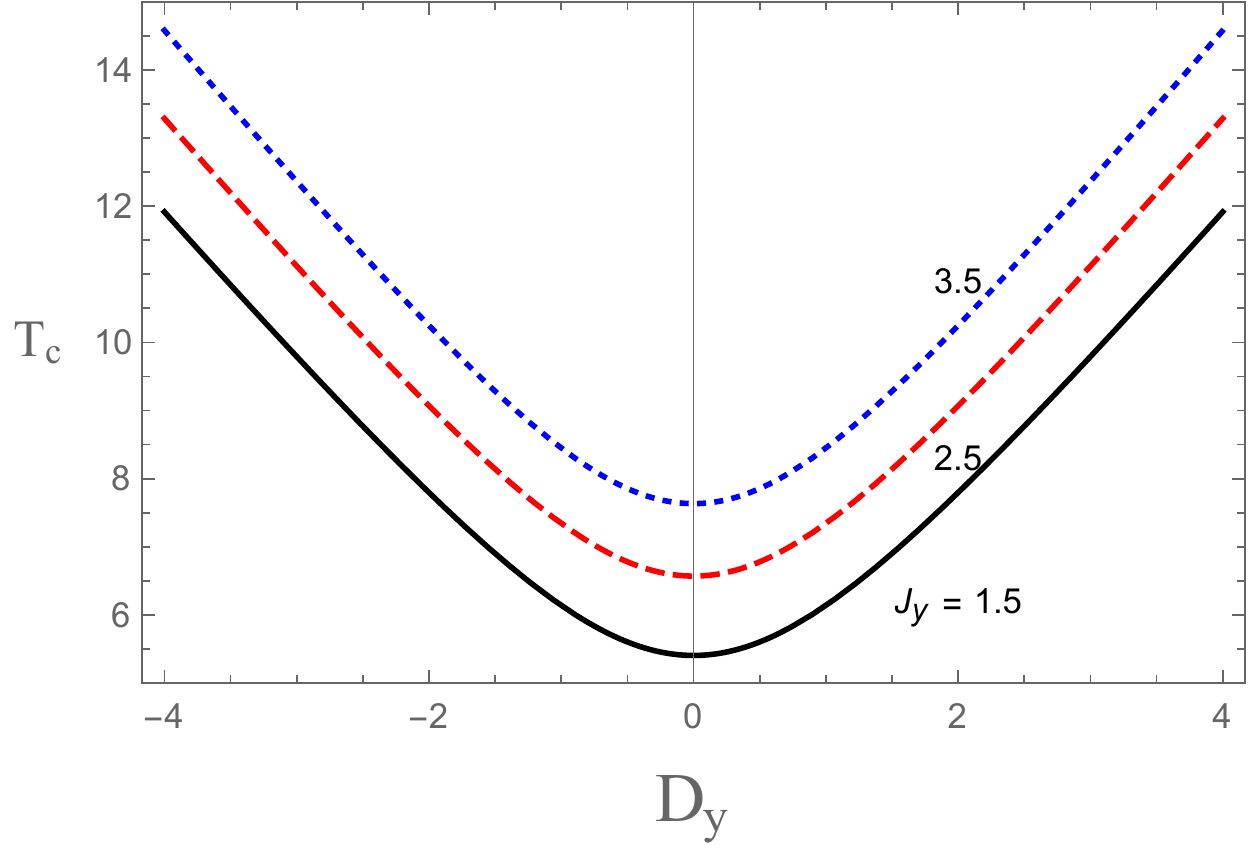} 
\includegraphics[height=5.0cm]{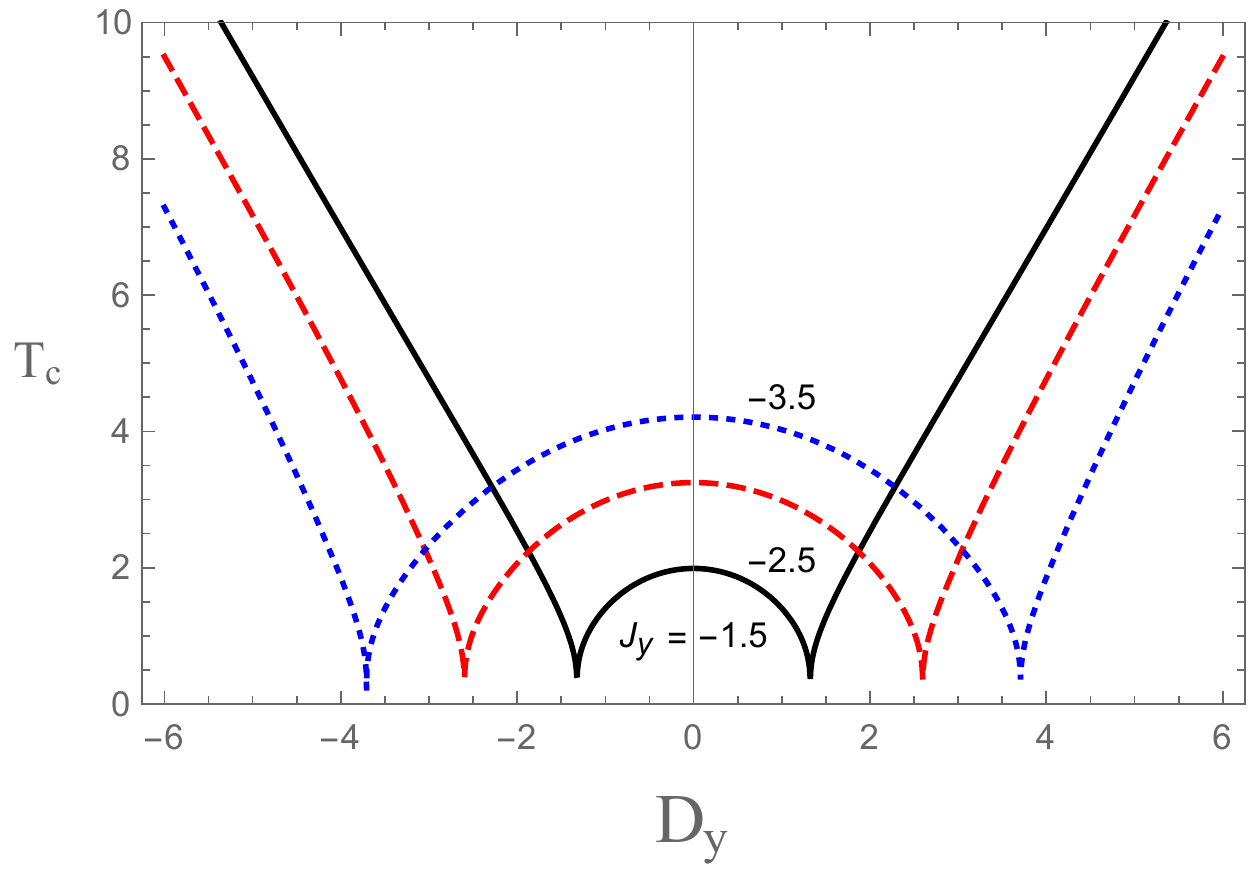}

\caption[fig5]{(Color online) The $D_y$-dependence of $T_c$ for antiferromagnetic (Fig. 5(a)) and ferromagnetic (Fig. 5(b)) cases with varying $J_y$.
(a) $J_x = 1$, $J_z = 2$, and $J_y = 1.5$ (black line), $2.5$ (red dashed line), $3.5$ (blue dotted line) are chosen. 
(b) $J_x = -1$, $J_z = -2$ and  $J_y = -1.5$ (black line), $- 2.5$ (red dashed line), $-3.5$ (blue dotted line) are chosen. 
}
\end{center}
\end{figure}

 For $X$ $Y$ $Z$ model with DM interaction Eq. (\ref{critical-3}) is replaced with 
\begin{eqnarray}
\label{critical-4}
&& e^{2 J_y / T_c} \frac{\sinh  (\eta / T_c )}{\cosh (|J_x - J_z| / T_c)} = 1   
                               \hspace{1.0cm} \mbox{when} \hspace{.3cm}  2 J_y + \eta \geq |J_x - J_z|                          \\   \nonumber
&& e^{-2 J_y / T_c} \frac{\sinh  (|J_x - J_z| / T_c )}{\cosh (\eta / T_c)} = 1  
                                                   \hspace{1.0cm} \mbox{when}  \hspace{.3cm} 2 J_y + \eta <  |J_x - J_z|.
\end{eqnarray}  
As in the case of $X$ $X$ $Z$ model the critical temperature $T_c$ for the antiferromagnetic case is determined by only first equation of Eq. (\ref{critical-4}). 
Thus, the behavior of ${\cal C} (\rho_Y)$ is similar to Fig. 4(a). Of course, both equations are used to determine $T_c$ for ferromagnetic case.
The boundary of two regions in Eq. (\ref{critical-4}) is  $D_y = D_{y,*}$, where 
\begin{equation}
\label{dystar}
D_{y,*} = \sqrt{(J_y - \tilde{J}_>) (J_y + \tilde{J}_<)}
\end{equation}
with $\tilde{J}_> = \max(J_x, J_z)$ and $\tilde{J}_< = \min(J_x, J_z)$. At this point $T_c$ becomes zero because $\eta_1 = \eta_3$ as 
Eq. (\ref{concur-y2}) shows. 

In Fig. 5
we plot the $D_y$-dependence of $T_c$ for antiferromagnetic (Fig. 5(a)) and ferromagnetic (Fig. 5(b)) cases with varying $J_y$. In Fig. 5(a) we choose
$J_x = 1$, $J_z = 2$, and $J_y = 1.5$ (black line), $2.5$ (red dashed line), $3.5$ (blue dotted line). Like $\rho_Z (T)$ $T_c$ increases with increasing $|D_y|$.
The critical temperature $T_c$ with fixed $D_y$ tends to increase with increasing $J_y$. In Fig. 5(b) we choose
$J_x = -1$, $J_z = -2$ and $J_y = -1.5$ (black line), $-2.5$ (red dashed line), $-3.5$ (blue dotted line). At the region $|D_y| \geq D_{y,*}$ 
$T_c$ increases with increasing $|D_y|$. With fixed $D_y$, in this region, $T_c$ tends to decreases with increasing $|J_y|$. 
At the region $|D_y| < D_{y,*}$ $T_c$ decreases with increasing $|D_y|$. With fixed $D_y$, $T_c$ increases with increasing $|J_y|$. 

For nether ferromagnetic and antiferromagnetic case the behavior of the concurrence is determined whether the second equation of Eq. (\ref{critical-4})
plays a role or not. Since it is similar to that of $D_x = D_y = 0$ case, we will not repeat the analysis. 

\section{Thermal Discord}
In this section we examine the temperature dependence of quantum discord\cite{discord1,discord2} for the thermal density matrices $\rho_Z (T)$ and $\rho_Y (T)$.

As well as quantum entanglement, quantum discord is another measure of quantum correlation for bipartite quantum system.
It is defined through discrepancy between two different quantum analogues of classical mutual information.  First analogue is 
\begin{equation}
\label{mutual1}
I (A:B) = S(A) + S(B) - S(A,B),
\end{equation}
where $S$ denotes a von Neumann entropy $S(\rho) = -\mbox{Tr} (\rho \log \rho)$. In our paper, all logarithms are taken to base $2$.
This is a quantum analogue of classical mutual information $I_{cl} (A:B) = H(A) + H(B) - H(A,B)$, where $H$ denotes a Shannon entropy. 
Another expression of classical mutual information is $I_{cl} (A:B) = H(A) - H(A|B) = H(B) - H(B|A)$, where 
$H(X|Y)$ is a conditional entropy of $X$ given $Y$. The quantum analogue of this representation\cite{discord1} is 
\begin{equation}
\label{mutual2}
J (A:B)_{\left\{ \Pi_j^B \right\}} = S(A) - \sum_j p_j S(A | \Pi_j^B ),
\end{equation}
where\footnote{Depending on the specific rules about the local operations and classical communication (LOCC) between parties A and B, 
one can define several different generalizations of $I_{cl} (A:B)$\cite{bro10,modi11}. Thus, several different quantum discords 
can be defined. Our definition (\ref{mutual2}) corresponds to the 
optimal efficiency of a one-way purification strategy\cite{bro10}.} $\left\{ \Pi_j^B \right\}$ denotes a 
complete set of measurement operators prepared by party $B$
and $S(A | \Pi_j^B )$ is a von Neumann entropy of party $A$ after party $B$ obtains a measurement outcome $j$. As obvious, 
$p_j$ is the probability of obtaining outcome $j$ in the quantum measurement. The general quantum mechanical postulates\cite{text} imply that
\begin{equation}
\label{mutual3}
p_j = \mbox{Tr}_{A,B} ( \Pi_j^B \rho_{AB} \Pi_j^B ),            \hspace{1.0cm}
S(A | \Pi_j^B ) = S \bigg( \rho \left(A | \Pi_j^B \right) \bigg),
\end{equation}
where $\rho \left(A | \Pi_j^B \right) = \mbox{Tr}_B ( \Pi_j^B \rho_{AB} \Pi_j^B ) / p_j$. Hence, unlike $I (A:B)$, $J (A:B)$ is dependent
on the complete set of measurement operators. The quantum discord is defined as 
\begin{equation}
\label{discord1}
{\cal D} (A:B) = \min \left[ I(A:B) - J(A:B)\right] = \min \left[S(B) - S(A,B) + \sum_j p_j S(A | \Pi_j^B ) \right]
\end{equation}
where the minimum is taken over all possible choice of the complete set of measurement operators\footnote{Although the authors in Ref. \cite{discord1}
consider
the projective measurement, the authors in Ref. \cite{discord2} consider the general measurement including positive operator-valued measure (POVM). Thus, the quantum discord in Ref. \cite{discord2} is the lower bound of that in \cite{discord1}.}.

\begin{figure}[ht!]
\begin{center}
\includegraphics[height=5.0cm]{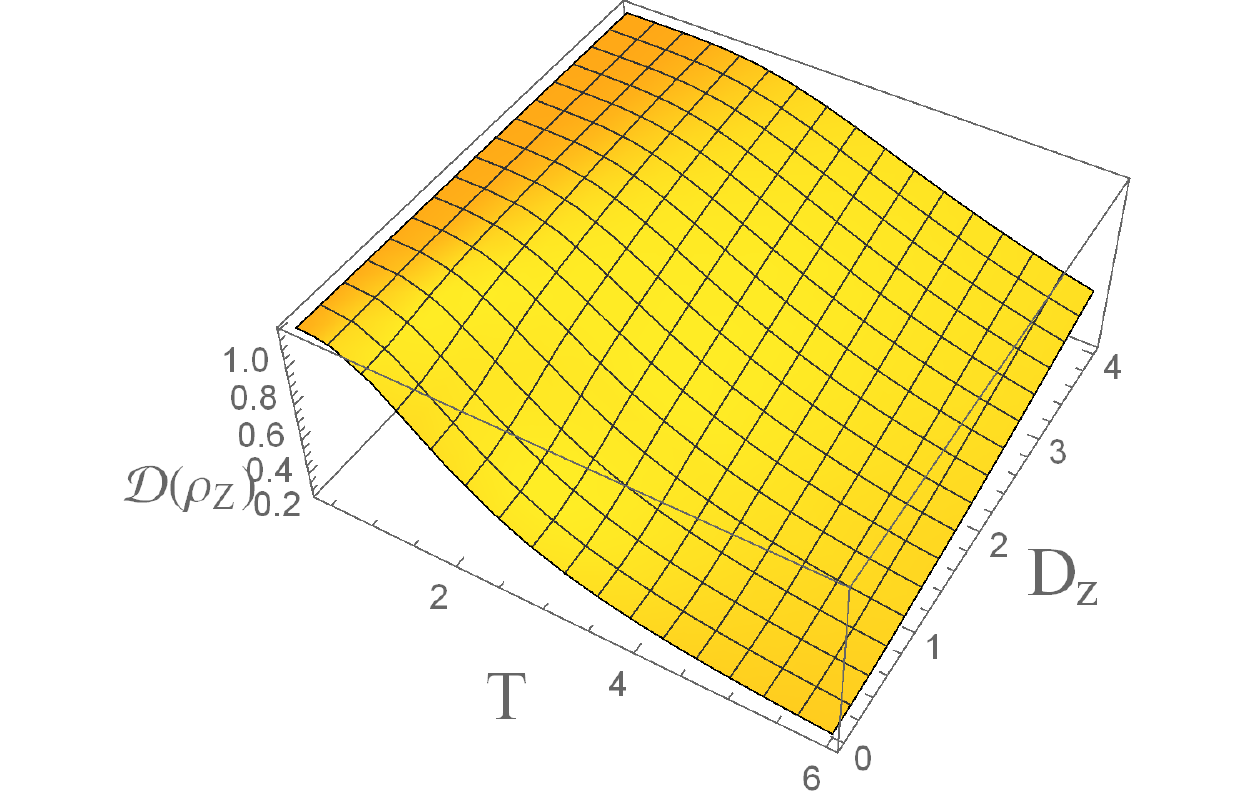} 
\includegraphics[height=5.0cm]{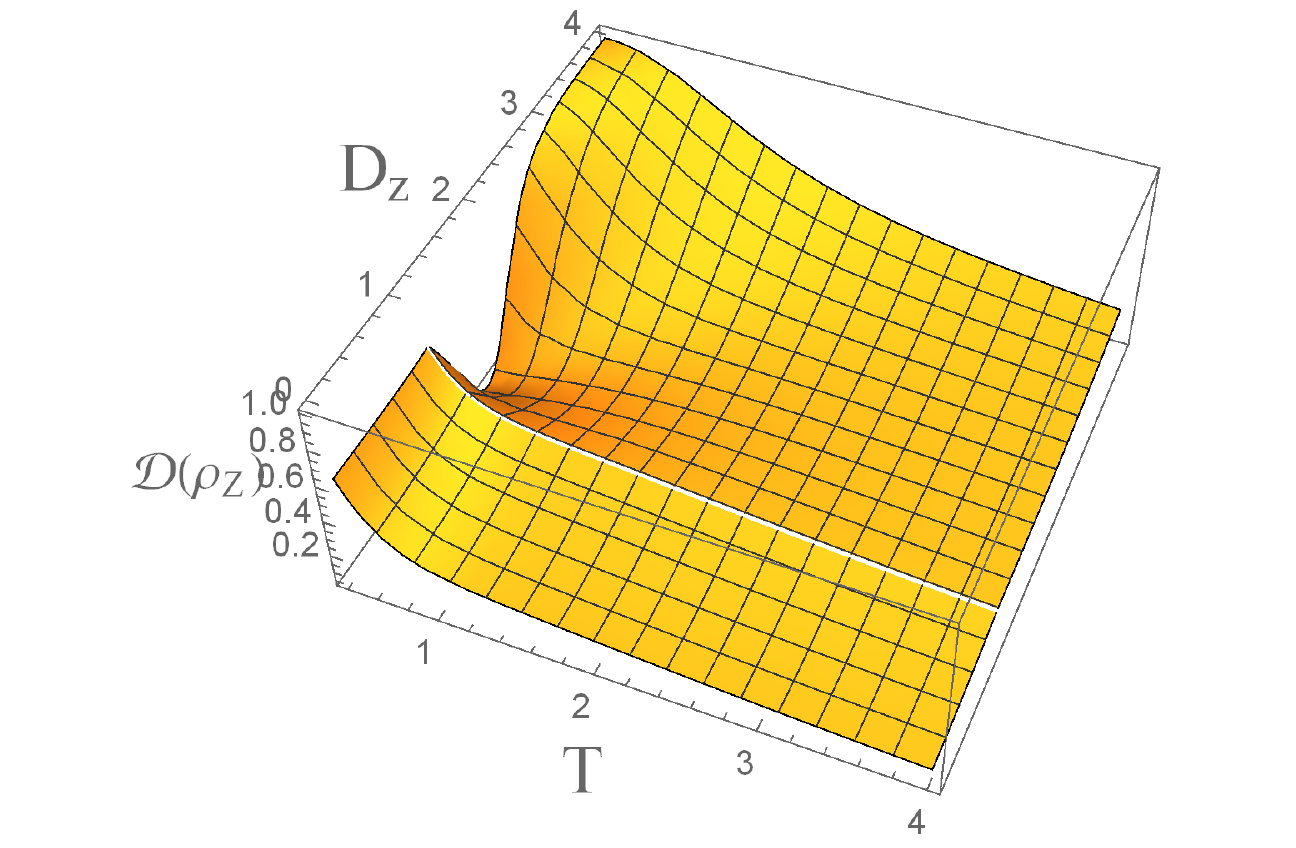}
\includegraphics[height=5.0cm]{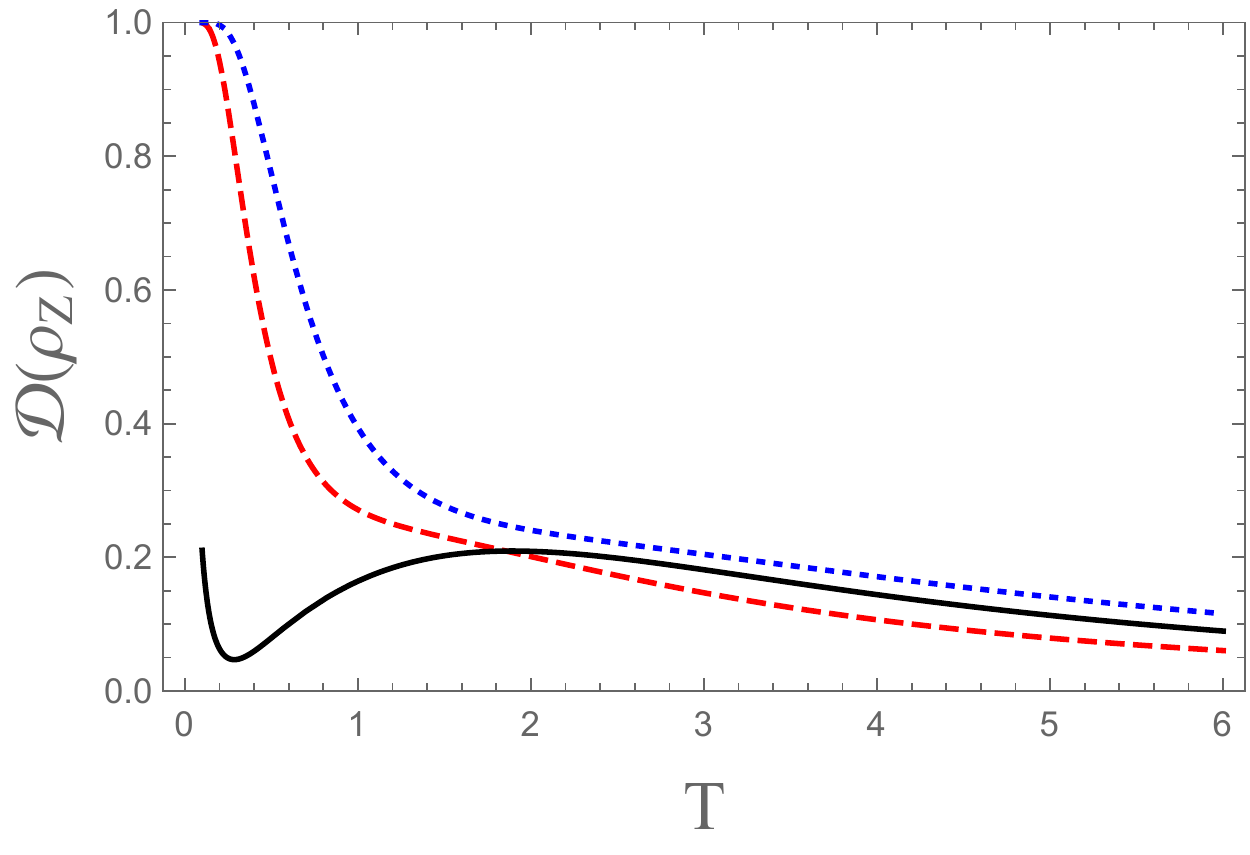}

\caption[fig6]{(Color online) The $T$- and $D_z$-dependence of the thermal discord ${\cal D} \left( \rho_Z \right)$ for (a) antiferromagnetic ($J_x = 1, J_y = 1.5, J_z = 2$) case and (b) ferromagnetic ($J_x = -1, J_y = -1.5, J_z = -2$ case. In order to examine the behavior of   ${\cal D} \left( \rho_Z \right)$ for 
ferromagnetic case at $D_z \approx \sqrt{7/2}$, ${\cal D} \left( \rho_Z \right)$ is plotted in Fig. 6(c)  with choosing $D_z = 0.8$ (red dashed line), $1.8$ (black line), $2.8$ (blue dotted line) without changing $J_{\alpha}$.
}
\end{center}
\end{figure}

The quantum discord for $\rho_Z (T)$ can be easily computed because it belongs to $X$-states. The quantum discord for general $X$-state was
computed in Ref.\cite{chen11,wang11}. For  $\rho_Z (T)$ the last term of Eq. (\ref{discord1}) becomes
\begin{equation}
\label{zdiscord-1}
\min  \sum_j p_j S(A | \Pi_j^B ) = \min \left( {\cal D}_{z,1},  {\cal D}_{z,2} \right)
\end{equation}
where
\begin{equation}
\label{zdiscord-2}
{\cal D}_{z,1} = -p \log p - (1-p) \log (1 - p)     \hspace{1.0cm}
{\cal D}_{z,2} = - 2 r \log r - 2 u \log u - 1
\end{equation}
with $p = \frac{1}{2} \left[ 1 + 2 (|s| + |v|) \right]$. The parameters $r$, $s$, $u$, and $v$ are defined in Eq. (\ref{density-z-boso}). Thus, quantum 
discord for  $\rho_Z (T)$ is given by 
\begin{eqnarray}
\label{zdiscord-3}
&& {\cal D} \left( \rho_Z \right) = 1 + (r + s) \log (r + s) + (r - s) \log (r - s)                         \\    \nonumber
&&\hspace{2.0cm} + (u + |v|) \log (u + |v|) + (u - |v|) \log (u - |v|) + \min \left( {\cal D}_{z,1},  {\cal D}_{z,2} \right).
\end{eqnarray}

In Fig. 6 we plot $T$- and $D_z$-dependence of the thermal discord ${\cal D} \left( \rho_Z \right)$ for (a) antiferromagnetic ($J_x = 1, J_y = 1.5, J_z = 2$) 
case and (b) ferromagnetic ($J_x = -1, J_y = -1.5, J_z = -2$ case. Although both thermal discords exhibit rapidly damping behavior with increasing $T$, unlike concurrence they do not reach to exact zero. This means that thermal discord does not vanish for separable states.  Furthermore, for the ferromagnetic case the 
thermal discord exhibits an extraordinary behavior at $D_z \approx D_{z,*} = \sqrt{7/2}$ like the concurrence. It seems to form a valley in this region. In order to examine 
this behavior more precisely we plot ${\cal D} \left( \rho_Z \right)$ with choosing $D_z = 0.8$ (red dashed line), $1.8$ (black line), $2.8$ (blue dotted line) in Fig. 6(c) without changing $J_{\alpha}$. This figure shows that ${\cal D} \left( \rho_Z \right)$ makes a local minimum at small $T$ region when 
$D_z = 1.8$ while other cases exhibit exponential damping behavior without local minimum. At $D_z = D_{z,*}$ and $T=0$ one can show $r = u = |v| = 1/4$
and $s = \mp 1/4$, which result in ${\cal D}_{z,1} = 0$ and ${\cal D}_{z,2} = 1$. Thus, the thermal discord of $\rho_Z (T)$ is exactly zero when 
$D_z = D_{z,*}$ and $T=0$. Since $T_c = 0$ at $D_z = D_{z,*}$, both thermal entanglement and thermal discord simultaneously vanish at this point. 

\begin{figure}[ht!]
\begin{center}
\includegraphics[height=5.0cm]{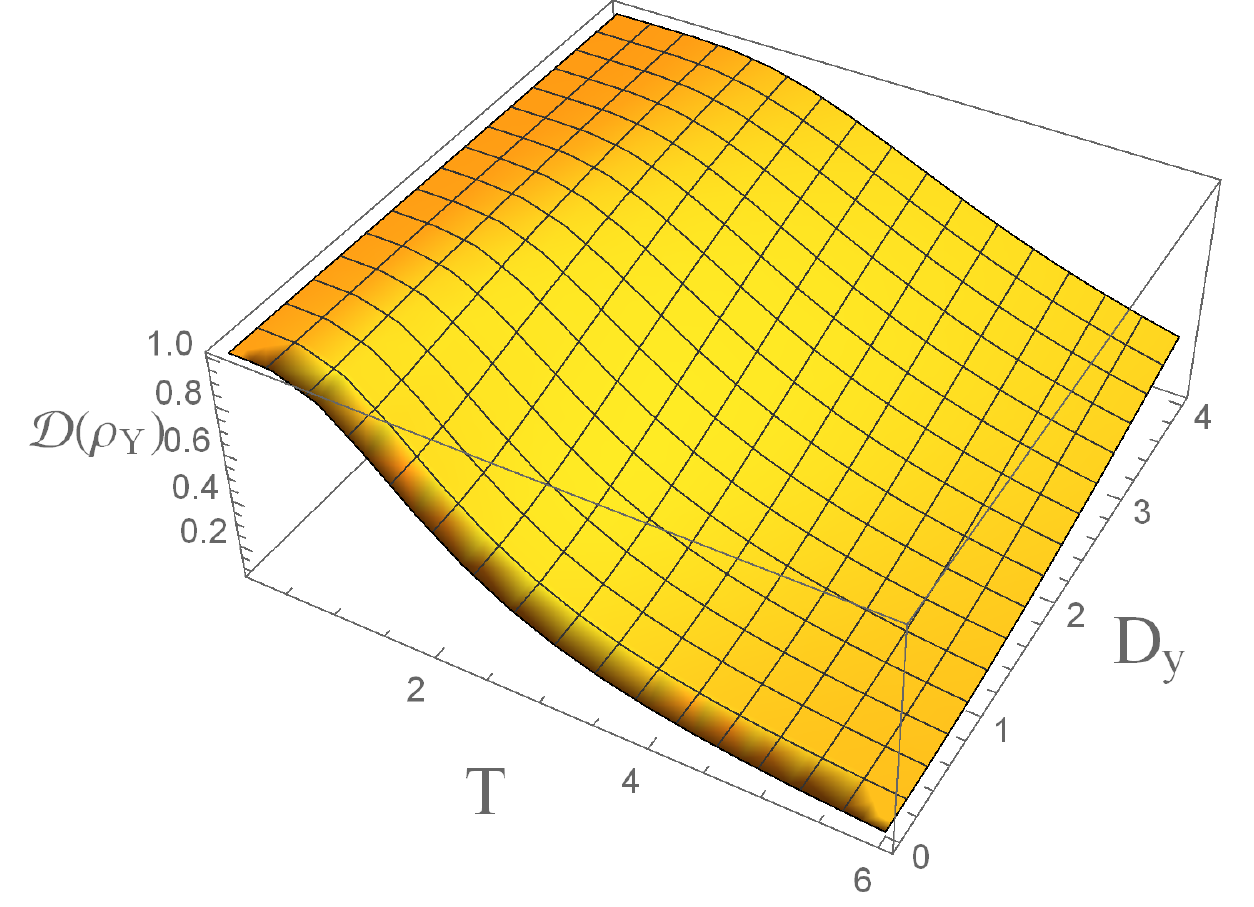} 
\includegraphics[height=5.0cm]{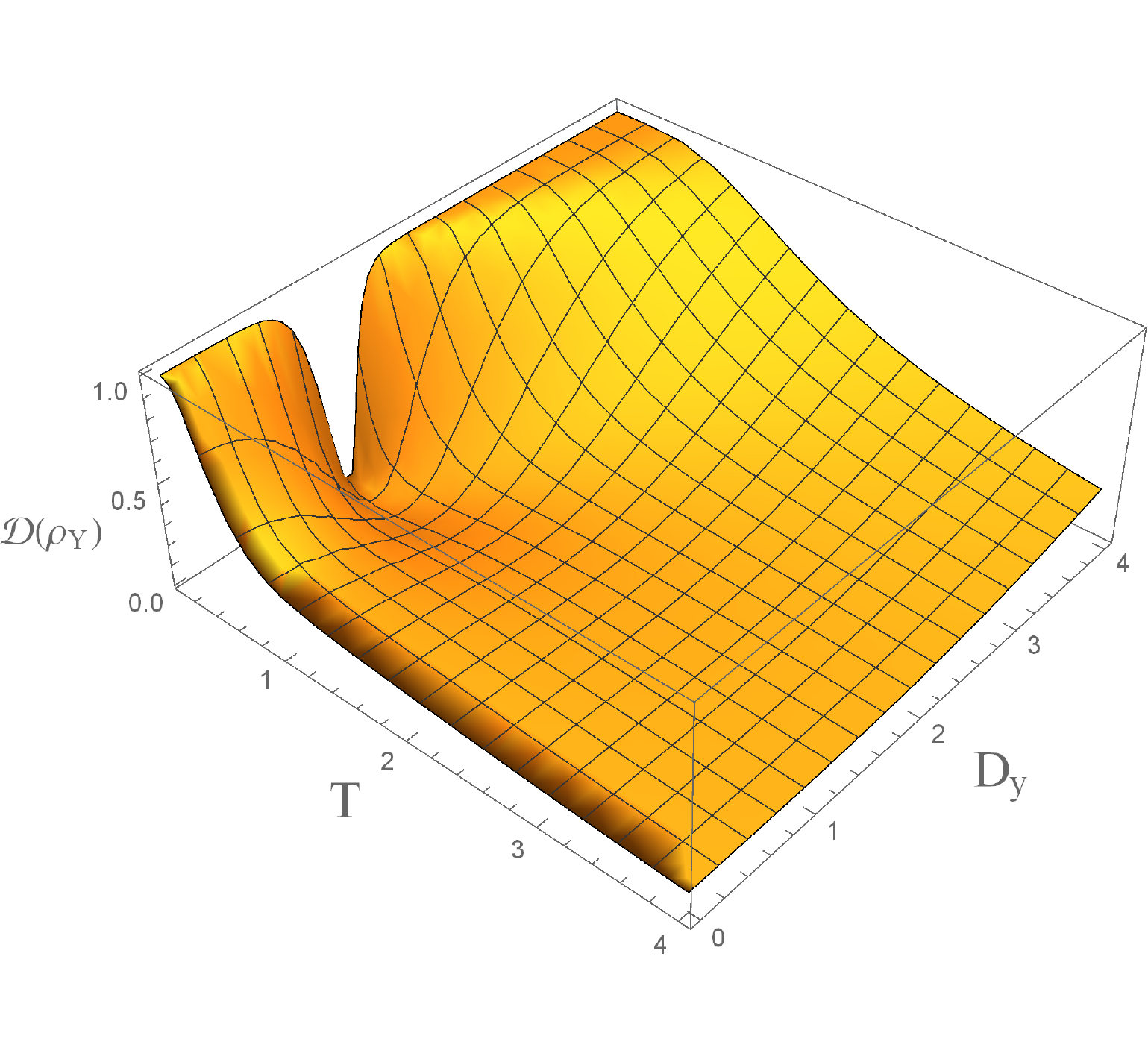}

\caption[fig7]{(Color online) The $T$- and $D_y$-dependence of the thermal discord ${\cal D} \left( \rho_Y \right)$ for (a) antiferromagnetic ($J_x = 1, J_y = 1.5, J_z = 2$) case and (b) ferromagnetic ($J_x = -1, J_y = -1.5, J_z = -2$) case.  For the ferromagnetic case the 
thermal discord exhibits an extraordinary behavior at $D_y \approx D_{y,*} = \sqrt{7} / 2 = 1.32$ like Fig. 6(b).
}
\end{center}
\end{figure}

Since $\rho_Y (T)$  does not belong to $X$-states, its thermal discord should be computed explicitly. This is carried out in appendix A, which gives
\begin{equation}
\label{ydiscord-1}
{\cal D} \left( \rho_Y \right) = 1 + \sum_{i = 1}^4 \frac{e^{-E_{i, y} / T}}{{\cal Z}_Y}  \log \left(  \frac{e^{-E_{i, y} / T}}{{\cal Z}_Y} \right) + 
h \left( \frac{1}{2} + \sqrt{y_{max}} \right)
\end{equation}
where $h(p) = -p \log p - (1 - p) \log (1-p)$ and 
\begin{equation}
\label{ydiscord-2}
y_{max} =  \frac{1}{2} \left[ 8 q^2 + (r_1 - u_1)^2 + (r_2 + u_2)^2 + |(r_1 - u_1) - (r_2 + u_2)| \sqrt{16 q^2 + [(r_1 - u_1) + (r_2 + u_2)]^2} \right].
\end{equation}

In Fig. 7 we plot $T$- and $D_y$-dependence of the thermal discord ${\cal D} \left( \rho_y \right)$ for (a) antiferromagnetic ($J_x = 1, J_y = 1.5, J_z = 2$) 
case and (b) ferromagnetic ($J_x = -1, J_y = -1.5, J_z = -2$ case. Like Fig. 6 both thermal discords do not reach to exact zero in spite of exponential damping 
with increasing $T$. For the ferromagnetic case the 
thermal discord exhibits an extraordinary behavior at $D_y \approx D_{y,*} = \sqrt{7} / 2 = 1.32$ like Fig. 6(b).

\begin{figure}[ht!]
\begin{center}
\includegraphics[height=5.0cm]{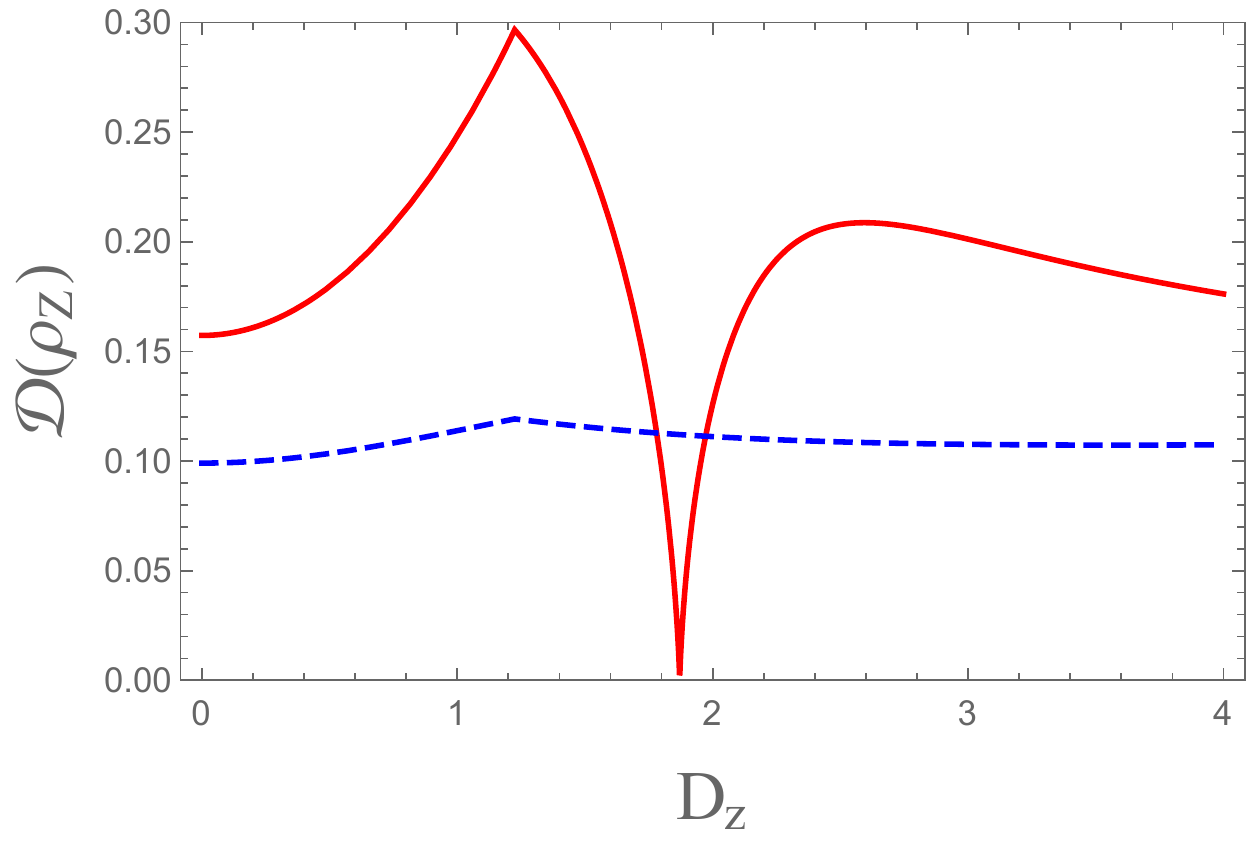} 
\includegraphics[height=5.0cm]{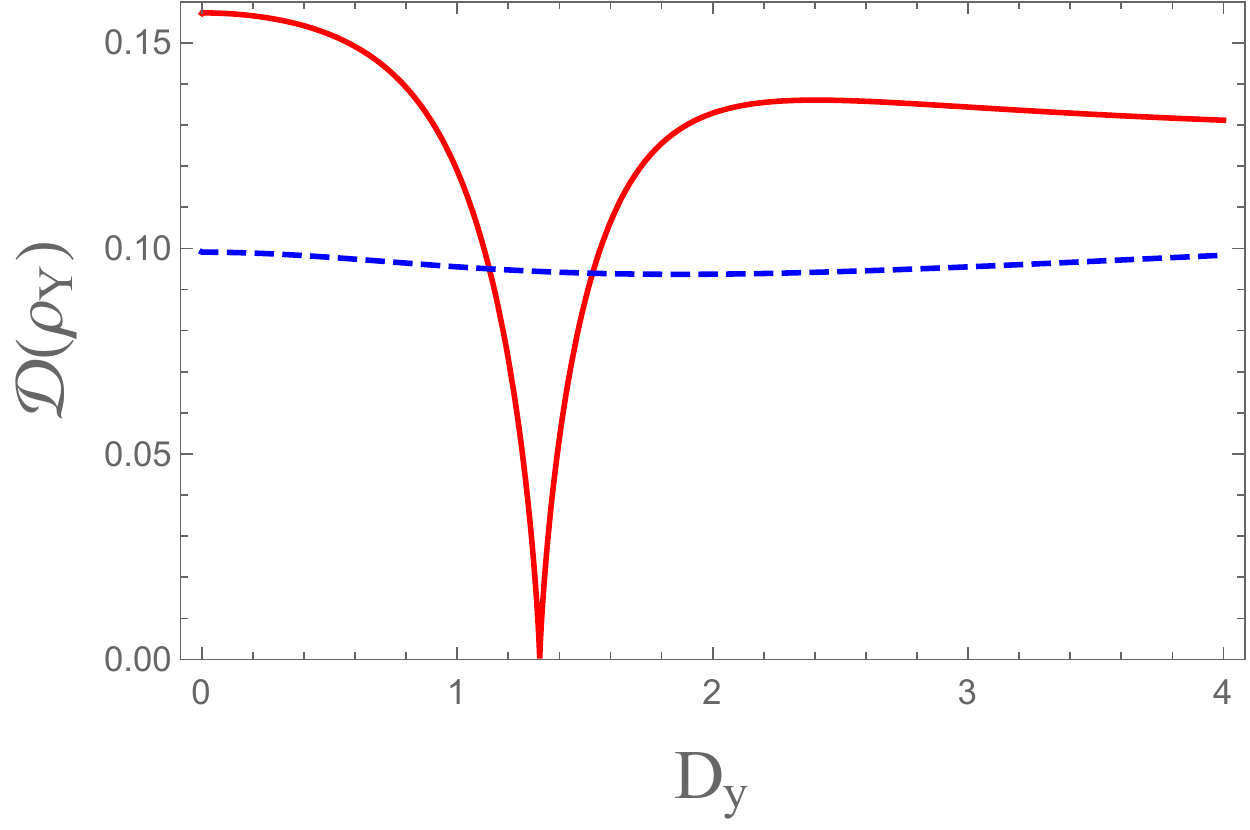}

\caption[fig8]{(Color online)  (a) The $D_z$-dependence of ${\cal D} (\rho_Z)$ and (b) $D_y$-dependence of ${\cal D} (\rho_Y)$ at $T= T_c$. 
In both figures red solid and blue 
dashed lines correspond to ferromagnetic  ($J_x = -1, J_y = -1.5, J_z = -2$) an antiferromagnetic  ($J_x = 1, J_y = 1.5, J_z = 2$) cases respectively. 
}
\end{center}
\end{figure}

Finally, we plot in Fig. 8 that (a) the $D_z$-dependence of ${\cal D} (\rho_Z)$ and (b) $D_y$-dependence of ${\cal D} (\rho_Y)$ at $T= T_c$. 
In both figures red solid and blue 
dashed lines correspond to ferromagnetic  ($J_x = -1, J_y = -1.5, J_z = -2$) and antiferromagnetic  ($J_x = 1, J_y = 1.5, J_z = 2$) cases, respectively. Thus, 
these figures show the discrepancy between thermal discord and thermal entanglement because concurrence at $T = T_c$ is exactly zero. Within the 
range $0 \leq D_z, D_y \leq 4$, these figures show  ${\cal D} (\rho_Z) \approx {\cal D} (\rho_Y) \approx 0.1$ for both antiferromagnetic cases. 
For ferromagnetic cases  ${\cal D} (\rho_Z)$ and ${\cal D} (\rho_Y)$ fall to zero at $D_z = \sqrt{7/2}$ and $\sqrt{7} / 2$ at $T = T_c$. 
This fact implies that for arbitrary ferromagnetic cases ($J_{\alpha} < 0$) both thermal discord and thermal entanglement simultaneously vanish at 
$D_z = D_{z,*}$ for $\rho_Z (T_c)$ and $D_y = D_{y,*}$ for $\rho_Y (T_c)$. If one extends the range of $D_z$ and $D_y$ in Fig. 8, both
antiferromagnetic and ferromagnetic ${\cal D} (\rho_Z)$ and ${\cal D} (\rho_Y)$ approach to one for $D_z, D_y \rightarrow \pm \infty$. Thus, maximal 
discrepancy between thermal discord and thermal entanglement occurs in this limit at the critical temperature.

\section{Conclusions}
We derive explicitly the thermal density matrices $\rho_Z (T)$ and $\rho_Y (T)$ for two-qubit Heisenberg $X$ $Y$ $Z$ chain with DM interaction
in the $z$- or $y$-direction. For each density matrix the thermal entanglement expressed by ${\cal C} (\rho_Z)$ or ${\cal C} (\rho_Y)$ is explicitly 
computed. Exploiting the explicit expressions of ${\cal C} (\rho_Z)$ or ${\cal C} (\rho_Y)$ we discuss on the quantum phase transition in detail. 
In particular, we focus on the critical temperature $T_c$, above which the thermal entanglement completely vanishes. This means that each density matrix
becomes separable state at $T \geq T_c$\footnote{Of course, this critical behavior of entanglement occurs when the DM interaction has arbitrary direction.
In order to show this fact more explicitly we discuss the critical behavior in appendix B when $D_x \neq 0$ and $D_y \neq 0$.}. 

For antiferromagnetic case ($J_{\alpha} > 0 \hspace{.2cm} \alpha = x, y, z$) the critical temperature is determined by single equation for each 
thermal density matrix. As a result, $T_c$ monotonically increases with increasing $|D_z|$ or $|D_y|$. 
For ferromagnetic case ($J_{\alpha} < 0 \hspace{.2cm} \alpha = x, y, z$), however, the situation is different. In this case the critical 
temperature is determined by two different equations in the two separate regions. For $\rho_Z (T)$, for example, these two separate regions are 
defined by $|D_z| < D_{z,*}$ and $|D_z| \geq D_{z,*}$.  Similarly, the two separate regions for $\rho_Y (T)$ are given by 
$|D_y| < D_{y,*}$ and $|D_y| \geq D_{y,*}$.
At the outer region $|D_z| \geq D_{z,*}$ or $|D_y| \geq D_{y,*}$ the critical 
temperature $T_c$ monotonically increases with increasing $|D_z|$ or $|D_y|$ like antiferromagnetic case.  At the inner region $|D_z| < D_{z,*}$ or 
$|D_y| < D_{y,*}$, however, $T_c$ decreases with increasing $|D_z|$ or $|D_y|$. As a result, the $D_{\alpha}$-  $\hspace{.1cm}(\alpha = z, y)$ and 
$T$-dependence of thermal entanglements for ferromagnetic case exhibit different behavior from those for antiferromagnetic case. 

Thermal discords ${\cal D} (\rho_Z)$ and ${\cal D} (\rho_Y)$ for $\rho_Z (T)$ and $\rho_Y (T)$ are explicitly derived. For antiferromagnetic case 
with fixed $D_z$ or $D_y$ both ${\cal D} (\rho_Z)$ and ${\cal D} (\rho_Y)$ exhibit  exponential damping behavior with increasing $T$, but they do not
reach to exact zero. For ferromagnetic case both ${\cal D} (\rho_Z)$ and ${\cal D} (\rho_Y)$ exhibit extraordinary behavior at 
$|D_z| \approx D_{z,*}$ and $|D_y| \approx D_{y,*}$. In these regions the $T$-dependence of  ${\cal D} (\rho_Z)$ or ${\cal D} (\rho_Y)$ involves
local minimum at small $T$ while in other region they exhibit exponential damping behavior without local minimum.  Although definitions of quantum entanglement and quantum discord are completely different, one can infer from our results that 
they exhibit similar behavior with each other because for ferromagnetic case both exhibit extraordinary behavior at the regions 
$|D_z| \approx D_{z,*}$ and $|D_y| \approx D_{y,*}$. At $D_z = D_{z,*}$ or $D_y = D_{y,*}$ both thermal entanglement and thermal discord for $\rho_Z (T)$ or $\rho_Y (T)$ simultaneously vanish at $T = T_c = 0$. 

One can apply our analysis when external magnetic field ${\bm B}$ is applied to our system\cite{li08}. In this case the Hamiltonian is changed into 
\begin{equation}
\label{final-3}
H_T = H + {\bm B} \cdot ({\bm \sigma}_1 + {\bm \sigma}_2 ),
\end{equation}
where $H$ is given in Eq. (\ref{hamil-total}). It seems to be of interest to examine the effect of external magnetic field and DM coupling constants in the
thermal entanglement and thermal discord. In particular,  it is of interest to analyze the behavior of thermal discord near the critical temperature in this model.

Another interesting issue is to examine the quantum phase transition by introducing $3$-spin Heisenberg model with DM interaction, whose Hamiltonian is
\begin{equation}
\label{3-qubit}
H_3 = \sum_{i=1}^{2} \left[ J_x \sigma_i^x \otimes \sigma_{i+1}^x  + J_y \sigma_i^y \otimes \sigma_{i+1}^y 
+ J_z \sigma_i^z \otimes \sigma_{i+1}^z 
 + {\bm D} \cdot \left( {\bm \sigma}_i \times {\bm \sigma}_{i+1} \right) \right].
\end{equation}
The tripartite entanglement was introduced in Ref.\cite{ckw,ou07-1}. It is of interest to compute the thermal tripartite entanglement and to discuss on the
quantum phase transition. It seems to be of interest to examine how the monogamy relation is changed with varying DM coupling constants and external temperature. 
  
One can extend our analysis to continuum system. For example, let us consider two coupled harmonic oscillator system, whose Hamiltonian is 
\begin{equation}
\label{final-4}
H = \frac{1}{2} \left( \dot{x}_1^2 + \dot{x}_2^2 \right) + \frac{1}{2} \left( \omega_1^2 x_1^2 + \omega_2^2 x_2^2 \right) - J x_1 x_2.
\end{equation}
In this case the thermal density matrix $\rho (x_1', x_2': x_1, x_2: T)$ can be derived exactly by making use of diagonalization of Hamiltonian and 
Euclidean path-integral technique\cite{feynman}. This thermal density matrix is generally mixed state. 
For mixed state entanglement is in general defined via a convex-roof method\cite{benn96,uhlmann99-1};
\begin{equation}
\label{final-5}
{\cal E} (\rho) = \min \sum_j p_j {\cal E} (\psi_j),
\end{equation}
where minimum is taken over all possible pure state decompositions, i.e. $\rho = \sum_j p_j \ket{\psi_j} \bra {\psi_j}$, with $0 \leq p_j \leq 1$. However,
we do not know how to derive the optimal decomposition for continuum thermal density matrix  $\rho (x_1', x_2': x_1, x_2: T)$. We would like to explore this
issue in the future.

{\bf Acknowledgement}:
This work was supported by the Kyungnam University Foundation Grant, 2017.

\newpage 

\begin{appendix}{\centerline{\bf Appendix A}}

\setcounter{equation}{0}
\renewcommand{\theequation}{A.\arabic{equation}}

In this appendix we compute the thermal discord for $\rho_Y (T)$ given in Eq. (\ref{density-y}). We define the measurement operators $\Pi_j^B$ as 
$\Pi_j^B = \ket{B_j} \bra{B_j} \hspace{.2cm} (j=1,2)$, where

\begin{equation}
\label{appen-1}
\ket{B_1} = \cos \frac{\theta}{2} \ket{0} + e^{i \phi} \sin \frac{\theta}{2} \ket{1}   \hspace{1.0cm}
\ket{B_2} = \sin \frac{\theta}{2} \ket{0} - e^{i \phi} \cos \frac{\theta}{2} \ket{1}.
\end{equation}
For our purpose, $\theta$ and $\phi$ are confined as $0 \leq \theta \leq \pi / 2$, $0 \leq \phi < 2 \pi$. 
 
Then, it is straightforward to show $G (\theta, \phi) = \sum_j p_j S(A | \Pi_j^B)$ becomes
\begin{equation}
\label{appen-2}
G (\theta, \phi) = - \left( \frac{1}{2} + \sqrt{y} \right) \log  \left( \frac{1}{2} + \sqrt{y} \right) -  \left( \frac{1}{2} - \sqrt{y} \right)
\log  \left( \frac{1}{2} - \sqrt{y} \right)
\end{equation}
where 
\begin{eqnarray}
\label{appen-3}
&&y = \left[ (r_1 - u_1)^2 + 4 q^2 \right] \cos^2 \theta + \left[ 4 q^2 \cos^2 \phi + r_2^2 + u_2^2 + 2 r_2 u_2 \cos 2\phi \right] \sin^2 \theta  
                                                                                                                                                                              \nonumber     \\
&&\hspace{4.0cm}  + 4 q \left[ (r_2 + u_2) - (r_1 - u_1) \right] \sin \theta \cos \theta \cos \phi.
\end{eqnarray}
The parameters $r_1$, $r_2$, $u_1$, $u_2$, and $q$ are explicitly given in Eq. (\ref{density-y-boso}). 

Now, we should minimize $G (\theta, \phi)$. We should note that $G (\theta, \phi)$ is a type of binary entropy function 
$h(p) = -p \log p -(1-p) \log (1 - p)$, which is concave with respect to $p$ and attains its maximum value of one at $p = 1/2$. Thus, 
minimizing $G (\theta, \phi)$ is exactly the same with maximizing $y$, i.e.,
\begin{equation}
\label{appen-4}
\frac{\partial y (\theta, \phi)}{\partial \theta} = 0    \hspace{2.0cm}  \frac{\partial y (\theta, \phi)}{\partial \phi} = 0.
\end{equation}
$ {\partial y (\theta, \phi)} /{\partial \phi} = 0$ yields three solutions
\begin{equation}
\label{appen-5}
\sin \theta = 0  \hspace{1.0cm}  \sin \phi = 0   \hspace{1.0cm} \tan \theta = \frac{q [(r_1 - u_1) - (r_2 + u_2)]}{2 \cos \phi (q^2 + r_2 u_2)},
\end{equation}
and $ {\partial y (\theta, \phi)} /{\partial \theta} = 0$ yields 
\begin{equation}
\label{appen-6}
\tan 2\theta = -\frac{4 q \cos \phi [(r_1 - u_1) - (r_2 + u_2)]}{4 (q^2 + r_2 u_2) \sin^2 \phi + (r_1 - u_1)^2 - (r_2 + u_2)^2}.
\end{equation}

The solution $\sin \theta = 0$ and Eq. (\ref{appen-6}) make $y$ to be 
\begin{equation}
\label{appen-7}
y_1 = (r_1 - u_1)^2 + 4 q^2.
\end{equation}
The solution $\sin \phi = 0$ and Eq. (\ref{appen-6}) generate two $y$ and larger one is 
\begin{equation}
\label{appen-8}
y_2 = \frac{1}{2} \left[ 8 q^2 + (r_1 - u_1)^2 + (r_2 + u_2)^2 + |(r_1 - u_1) - (r_2 + u_2)| \sqrt{16 q^2 + [(r_1 - u_1) + (r_2 + u_2)]^2} \right].
\end{equation}
In order for the third solution of Eq. (\ref{appen-5}) and Eq. (\ref{appen-6}) to be consistent we need a condition
$$4 \left(q^2 + r_2 u_2 \right)^2 + [(r_1 - u_1) - (r_2 + u_2)] \left[ 2 q^2 (r_2 + u_2) + r_2 u_2 \left\{ (r_1 - u_1) + (r_2 + u_2) \right\} \right] = 0$$ 
identically because $\tan 2 \theta = 2 \tan \theta / (1 - \tan^2 \theta)$. However, one can show easily that this condition does not hold identically 
by making use of Eq. (\ref{density-y-boso}). There is another possibility. The boundary value of $y$ can be maximum although it is not 
local maximum. Thus, another candidate for $\max y$ is $y$ at $\theta = \pi / 2$, $\phi = 0$, which is 
\begin{equation}
\label{appen-9}
y_3 = (r_2 + u_2)^2 + 4 q^2.
\end{equation}
It is easy to show $y_{max} = \max (y_1, y_2, y_3) = y_2$. Thus, $\min G (\theta, \phi)$ becomes
\begin{equation}
\label{appen-9}
\min G (\theta, \phi) = - \left( \frac{1}{2} + \sqrt{y_2} \right) \log  \left( \frac{1}{2} + \sqrt{y_2} \right) -  \left( \frac{1}{2} - \sqrt{y_2} \right)
\log  \left( \frac{1}{2} - \sqrt{y_2} \right).
\end{equation}

\end{appendix}
 \newpage 

\begin{appendix}{\centerline{\bf Appendix B}}

\setcounter{equation}{0}
\renewcommand{\theequation}{B.\arabic{equation}}

In this appendix we will show that the ESD phenomenon of entanglement still occurs when multiple DM components are present. For simplicity, we choose 
$J_x = J_y \equiv J$ and $D_z = 0$. In this case the Hamiltonian becomes
\begin{eqnarray}
\label{b-hamil}
H_{XY} = \left(         \begin{array}{cccc}
                      J_z  & i D_x + D_y  & -i D_x - D_y  &  0                 \\
                      -i D_x + D_y  &  -J_z  &  2 J  &  i D_x + D_y           \\
                      i D_x - D_y  &  2 J  &  -J_z  &  -i D_x - D_y           \\
                      0  &  -i D_x + D_y  &  i D_x - D_y  &  J_z
                              \end{array}                                      \right).
\end{eqnarray}
The eigenvalues and corresponding eigenvectors of $H_{XY}$ are summarized in Table III. In this Table $\zeta$, ${\cal N}_3$ and ${\cal N}_4$ are given by
\begin{eqnarray}
\label{b-table3}
&&\hspace{2.0cm} \zeta = \sqrt{4(D_x^2 + D_y^2) + (J + J_z)^2}   \\   \nonumber
&&  {\cal N}_3^2 = 4 \zeta \left\{\zeta - (J + J_z) \right\}   \hspace{1.0cm}
 {\cal N}_4^2 = 4 \zeta \left\{\zeta + (J + J_z) \right\}.
\end{eqnarray}

\begin{center}
\begin{tabular}{c|c} \hline \hline
eigenvalues of $H_{XY}$ & corresponding eigenvectors                                     \\  \hline \hline
$E_{1} = 2 J - J_z$ &  $\ket{E_1} = \frac{1}{\sqrt{2}} \left(\ket{01} + \ket{10} \right)$                    \\    
$E_{2} = J_z$ & $\ket{E_2} = \frac{1}{\sqrt{2}} \left( \sqrt{ \frac{D_x - i D_y}{D_x + D_y}}\ket{00} + \sqrt{ \frac{D_x + i D_y}{D_x - i D_y}} \ket{11}              
                                                                                                                                                                                \right)$                    \\  
$E_{3} = - J + \zeta$  &  $\ket{E_3} = 
\frac{1}{{\cal N}_3}  \bigg[ 2 (- D_x + i D_y) \ket{00} -i (J + J_z - \zeta) (\ket{01} - \ket{10}) + 2 (D_x + i D_y) \ket{11} \bigg]$                \\
$E_{4} = - J - \zeta$  &  $\ket{E_4} = 
\frac{1}{{\cal N}_4}  \bigg[ 2 (- D_x + i D_y) \ket{00} -i (J + J_z + \zeta) (\ket{01} - \ket{10}) + 2 (D_x + i D_y) \ket{11} \bigg]$       \\ \hline
\end{tabular}

\vspace{0.3cm}
Table I: Eigenvalues and eigenvectors of $H_{XY}$ 
\end{center}
\vspace{0.5cm}

\begin{figure}[ht!]
\begin{center}
\includegraphics[height=5.0cm]{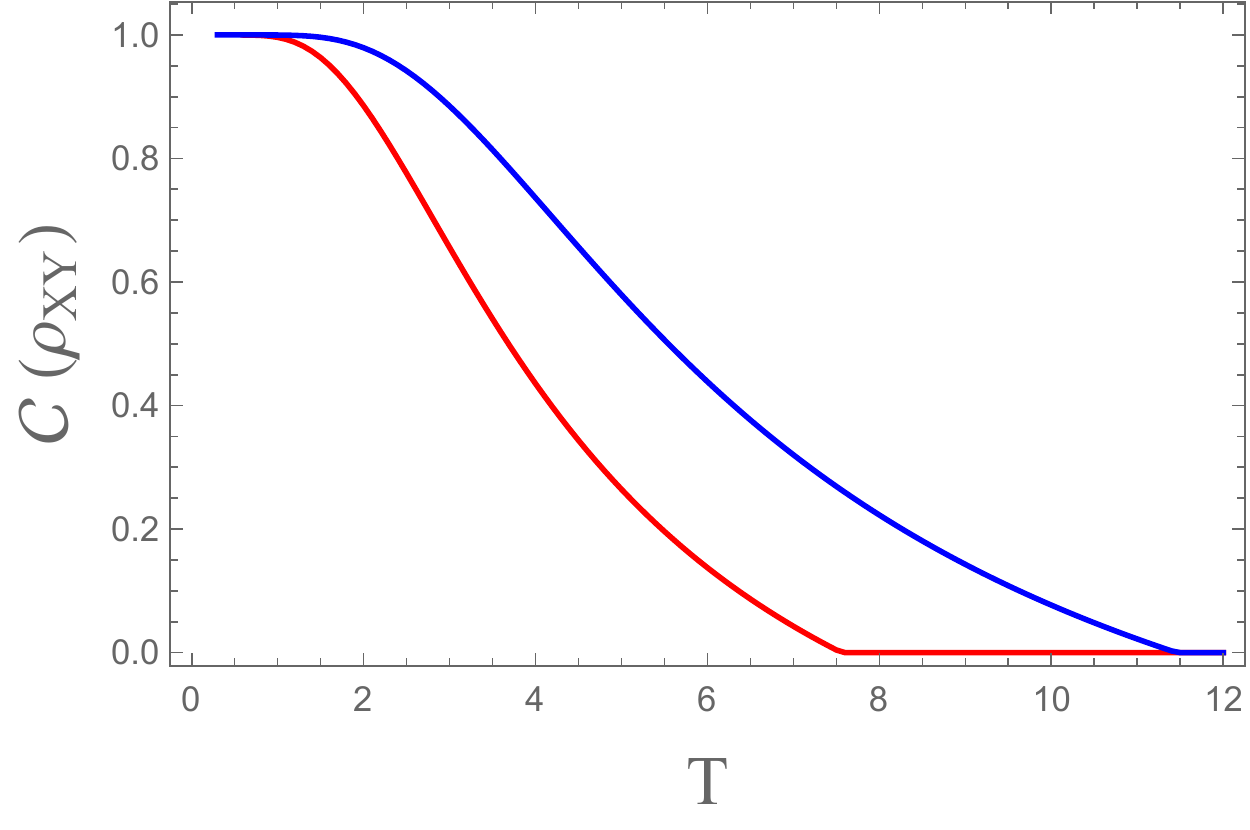} 

\caption[fig9]{(Color online) The $T$-dependence of concurrence with  $D = \sqrt{5}$ (red line)  and  $D = \sqrt{17}$ (blue line) when $J=1$ and $J_z = 2$. As expected, both decrease with increasing $T$, and eventually go to zero. The critical temperature are  $T_c \sim 7.6$ for  $ D = \sqrt{5}$ and $T_c \sim 11.5$ for  $D = \sqrt{17}$ approximately. 
   }
\end{center}
\end{figure}

Thus, the spectral decomposition of $H_{XY}$ can be written as 
\begin{equation}
\label{boptimal-xy}
H_{XY} = \sum_{i=1}^4 E_{i} \ket{E_i} \bra{E_i}.
\end{equation}

Then the partition of this system is 
\begin{equation}
\label{b-partition}
{\cal Z}_{XY} = \mbox{Tr} \left[e^{-\beta H_{XY}} \right] = 2 e^{-\beta J} \cosh [ \beta (J - J_z)] + 2 e^{\beta J} \cosh (\beta \zeta).
\end{equation}
Then the thermal density matrix in this case becomes
\begin{eqnarray}
\label{b-thermal}
\rho_{XY} (T) =  \left(                      \begin{array}{cccc}
                                                                      z_1  &  z_3  &  -z_3  &  z_4                      \\
                                                                      z_3^*  &  z_2  &  z_5  &  z_3                     \\
                                                                      -z_3^*  &  z_5  &  z_2  &  -z_3                   \\
                                                                      z_4^*  &  z_3^*  &  -z_3^*  &  z_1
                                                                                            \end{array}                                \right)
\end{eqnarray}
where
\begin{eqnarray}
\label{b-thermal-boso}
&& z_1 = \frac{1}{2 {\cal Z}_{XY}} \left[ e^{-\beta J_z} + \frac{e^{\beta J}}{\zeta} \left\{ \zeta \cosh (\beta \zeta) - (J + J_z) \sinh (\beta \zeta) \right\}  \right]                                          \\    \nonumber
&& z_2 = \frac{1}{2 {\cal Z}_{XY}} \left[ e^{-\beta (2 J - J_z)} + \frac{e^{\beta J}}{\zeta} \left\{ \zeta \cosh (\beta \zeta) + (J + J_z) \sinh (\beta \zeta) \right\}  \right]                               \\    \nonumber
&&z_3 =  \frac{1}{ {\cal Z}_{XY}} \frac{-i (D_x - i D_y)}{\zeta} e^{\beta J} \sinh (\beta \zeta)                   \\     \nonumber
&&z_4 =  \frac{1}{2 {\cal Z}_{XY}} \frac{D_x - i D_y}{D_x + i D_y}
\left[ e^{-\beta J_z} - \frac{e^{\beta J}}{\zeta} \left\{ \zeta \cosh (\beta \zeta) - (J + J_z) \sinh (\beta \zeta) \right\}  \right]  
                                                                                                                                                                                      \\   \nonumber
&&z_5 =  \frac{1}{2 {\cal Z}_{XY}} \left[ e^{-\beta (2 J - J_z)} - \frac{e^{\beta J}}{\zeta} \left\{ \zeta \cosh (\beta \zeta) + (J + J_z) \sinh (\beta \zeta) \right\}  \right].
\end{eqnarray}

In order to compute the concurrence we should compute the eigenvalues of $R = \rho_{XY} (\sigma_y \otimes \sigma_y) \rho_{XY}^*  (\sigma_y \otimes \sigma_y)$. One eigenvalue is $(z_2 + z_5)^2$ and the remaining three eigenvalues are roots of the following cubic equation:
\begin{equation}
\label{b-cubic}
\Lambda^3 - \alpha_1 \Lambda^2 + \alpha_2 \Lambda - \alpha_3 = 0
\end{equation}
where 
\begin{eqnarray}
\label{b-cubic-1}
&&\alpha_1 = 2 z_1^2 + 8 a_1 + 2 a_2 + a_4^2                               \\     \nonumber
&&\alpha_2 = z_1^4 + (4 a_1 + a_2)^2 - 8 z_1 a_3 - 4 a_3 a_4 + 8 z_1 a_1 (z_1 - a_4) - 2 a_2 (z_1^2 - a_4^2) + 2 z_1^2 a_4^2      \\    \nonumber
&&\alpha_3 = \left[ 2 a_3 - 4 z_1 a_1 + (z_1^2 - a_2) a_4 \right]^2
\end{eqnarray}
with 
\begin{equation}
\label{b-cubic-2}
a_1 = |z_3|^2  \hspace{.5cm} a_2 = |z_4|^2  \hspace{.5cm}  a_3 = z_3^2 z_4^* + (z_3^*)^2 z_4  \hspace{.5cm}  a_4 = z_2 - z_5.
\end{equation}
It is worthwhile noting that the $D_x$- and $D_y$-dependence of the eigenvalues are only via $D = \sqrt{D_x^2 + D_y^2}$. Thus the concurrence ${\cal C} (\rho_{XY})$ has rotation  symmetry in $(D_x, D_y)$-plane. 

Solving the cubic equation (\ref{b-cubic}) on the analytical ground  is very cumbersome work, because the roots have too long expressions.  
Thus, we rely on the numerical method from present stage. The $T$-dependence of  ${\cal C} (\rho_{XY})$ with  $(D_x = 1, D_y = 2)$ and  $(D_x = 1, D_y = 4)$ is plotted in Fig. 9 as red and blue lines respectively when $J=1$ and $J_z = 2$. As expected, both decrease with increasing $T$, and eventually go to zero. The critical temperature are  $T_c \sim 7.6$ for  $(D_x = 1, D_y = 2)$ and $T_c \sim 11.5$ for  $(D_x = 1, D_y = 4)$ approximately.

\end{appendix}

\end{document}